\documentclass[12pt]{article}
\pdfoutput=1
\usepackage{jheppub}
\usepackage{amsmath,amssymb}
\usepackage{graphicx}

\makeatletter

\@addtoreset{equation}{section}
\makeatother

\newcommand{\beq}{\begin{equation}}
\newcommand{\eeq}{\end{equation}}

\newcommand{\bC}{\ensuremath{\mathbb{C}}}

\newcommand{\bH}{\ensuremath{\mathbb{H}}}

\newcommand{\bP}{\ensuremath{\mathbb{P}}}

\newcommand{\bR}{\ensuremath{\mathbb{R}}}

\newcommand{\bZ}{\ensuremath{\mathbb{Z}}}

\newcommand{\scN}{\ensuremath{\mathcal{N}}}

\newcommand{\scS}{\ensuremath{\mathcal{S}}}

\newcommand{\scZ}{\ensuremath{\mathcal{Z}}}

\begin{document}

\baselineskip=18pt  
\baselineskip 0.7cm

\begin{titlepage}

\setcounter{page}{0}

\renewcommand{\thefootnote}{\fnsymbol{footnote}}

\begin{flushright}
\end{flushright}

\vskip 1.5cm

\begin{center}
{\LARGE \bf
Superconformal Partition Functions
and Non-perturbative Topological Strings

\vskip 1.5cm 
}

{\large
Guglielmo Lockhart and Cumrun Vafa
\\
\medskip
}

\vskip 0.5cm

{
\small\it
Jefferson Physical Laboratory, Harvard University, Cambridge, MA 02138, USA
\\
\medskip
}
\end{center}

\centerline{{\bf Abstract}}
\medskip
\noindent

We propose a non-perturbative definition for refined topological strings.  This can be used
to compute the partition function of superconformal theories in 5 dimensions on squashed $S^5$ and
the superconformal index of a large number of 6 dimensional  $(2,0)$ and $(1,0)$ theories, including that of $N$ coincident M5 branes.
The result can be expressed as an integral over the product of three combinations of topological
string amplitudes.  $SL(3,{\bf Z})$  modular transformations acting by inverting the coupling
constants of the refined topological string play a key role.

\end{titlepage}
\setcounter{page}{1} 

\section{Introduction}

Topological strings have been defined perturbatively, but it is certainly interesting to ask whether one can find a non-perturbative definition for them.  In a strong sense topological strings,
which capture the BPS content of the deformations of the superconformal theories,
 compute relevant amplitudes for supersymmetric partition functions
of superconformal theories.  Thus one idea is to reverse the statement and
define non-perturbative topological strings using supersymmetric partition functions.

The relation between topological string partition functions and superconformal index for ${\cal N}=1$ 5d theories has
been explored in \cite{Vafa:2012fi,Iqbal:2012}.  The aim of this paper is to extend
this relation in two directions:  Given the relation between
superconformal partition functions and topological strings we come up with both
a definition of non-perturbative topological strings on the one hand, and also a proposal
for how to use topological strings to compute certain supersymmetric partition functions.  In
particular, we focus on the partition function of ${\cal N}=1$ superconformal theories in 5d on $S^5$ and
superconformal ${\cal N}=(2,0)$ and ${\cal N}=(1,0)$ theories in 6d on $S^5\times S^1$.   The perturbative
parts of the superconformal partition functions were computed for certain gauge theories
on $S^5$ \cite{Kallen:2012cs, Kallen:2012va,Kim:2012av, Kallen:2012zn, Jafferis:2012iv}, and using this ingredient and the condition that the BPS content captured by topological
strings behaves as the fundamental degrees of freedom of the theory, an idea advanced
in \cite{Iqbal:2012}, we propose not
only a way to compute
the full answer for superconformal partition functions on $S^5$, but also a non-perturbative definition for topological strings.  Moreover by
viewing 6d $(2,0)$ and $(1,0)$ superconformal theories compactified on $S^1$ as
a supersymmetric system in 5d, we are able to also compute the superconformal index for
a large class of $(2,0)$ (and in particular N coincident M5 branes) and $(1,0)$ theories in 6 dimensions.

The highly non-trivial aspect of this proposal is that the full non-perturbative aspect of the
topological partition function 
enters because we have coupling constants of topological strings inverted.  In particular, roughly speaking
the proposal for the non-perturbative topological string
partition function $Z_{np}$ takes the form (which  will be made more precise later in the paper)\footnote{
By analytic continuation this can also be written in the form $Z_{np}=Z^{top}\cdot Z^{top}\cdot Z^{top}$.}
$$Z_{np}(t_i,m_j,\tau_1,\tau_2)={Z^{top}(t_i,m_j;\tau_1,\tau_2)\over Z^{top}(t_i/\tau_1,m_j/\tau_1; -1/\tau_1,\tau_2/\tau_1)\cdot Z^{top}(t_i/\tau_2,m_j/\tau_2; \tau_1/\tau_2,-1/\tau_2)}$$
where $t_i, m_j$ are normalizable and non-normalizable Kahler classes, and $\tau_1,\tau_2$
are the two couplings of the refined topological strings.
Of course to define exactly what  this means we have to be more precise and we
 use the BPS degeneracies captured by topological strings to give a precise meaning
to $Z_{np}$.  Furthermore,  the superconformal partition function on $S^5$ is written in terms of this composite
non-perturbative $Z$ by
$$Z_{S^5}(m_j;\tau_1,\tau_2)=\int dt_i \ Z_{np}(t_i,m_j;\tau_1,\tau_2)$$
where $m_j$ are interpreted as mass parameters and $\tau_1,\tau_2$ can
be viewed as squashing parameters for $S^5$.  The relevant 5d theories we consider
can be viewed as compactification of M-theory on singular loci of Calabi-Yau manifolds where some 4-cycles have shrunk \cite{Witten:1996qb, Seiberg:1996bd, Morrison:1996xf, Douglas:1996xp, Intriligator:1997pq}.  For a subset of these, which geometrically engineer a gauge theory \cite{Katz}, $Z^{top}$ can
be identified with the 5d gauge theory partition function \cite{Nekrasov:2003rj}, with $\tau_i=\epsilon_i$.
We can also consider 6d superconformal theories:  There are two classes of them,
with $(2,0)$ or $(1,0)$ supersymmetry.   A large class of these theories can be obtained as
F-theory on elliptic 3-folds (in the case of  $(2,0)$ it corresponds
to a constant elliptic fiber).  Compactifying these theories on a circle
down to 5 dimensions leads to dual descriptions involving M-theory
on elliptic Calabi-Yau threefolds.  Upon further compactification on $S^5$,
we can use the resulting non-perturbative topological string on elliptic Calabi-Yau
threefold to compute the partition function on $S^5$.  This leads
to the partition function of the 6d theory on $S^1\times S^5$, i.e.  it leads
to the computation of the 6d superconformal index, where $m_j$ correspond
to fugacities for flavor symmetry and $\tau_{1,2}$ correspond to parameters of supersymmetric
rotations on $S^5$.  Moreover one of the fugacities $m_i$ corresponds to the Kahler class
$\tau$ of the elliptic fiber.  This will correspond to the extra parameter in the superconformal
$(1,0)$ theory.  For the $(2,0)$ theory the superconformal
index depends on 4 parameters.  In this case the corresponding topological theory
computes the partition function of $\scN=2^*$ gauge theory in 
5 dimensions and the mass and coupling constant of the gauge theory
correspond to the two additional parameters
of the $(2,0)$ 6d index (see \cite{Kim:2012av} for a related
discussion).  Thus, we are able to compute the
superconformal index for ${\cal N}=(2,0)$ systems in 6 dimensions.  

We can also consider Lagrangian defects of topological strings.  These
lead to 3d theories living on the non-compact part of the M5 brane wrapping
the Lagrangians.  Upon compactification on $S^3$ these can also be viewed
as a non-perturbative completion of the open topological string, which has already
been considered in \cite{Cecotti:2010fi, Cecotti:2011iy, Pasquetti:2011fj,dimofte}, 
In particular the structure for the open string part has the form
$$Z_{np}^{open}(\dots)={Z^{open}(t_i,m_j,x_k; \tau)\over Z^{open}(t_i/\tau,m_j/\tau,x_k/\tau; -1/\tau)},$$
where the $t_i,m_i$ are closed string parameters and $x_k$ label open string moduli.
The corresponding partition function on $S^3$ is given by
$$Z_{S^3}=\int dx_k \ Z_{np}^{open}(t_i,m_j,x_k;\tau).$$

The organization of this paper is as follows:  In section 2 we review the relation between
open topological strings and the $S^3$ partition function of M5 branes wrapping Lagrangians
in CY.  In section 3 we study the partition function of the ${\cal N}=1$ superconformal theories
in 5 dimensions.  In section 4 we propose a non-perturbative definition of
topological strings which can be used for the computation of these amplitudes. In section 5, we offer a possible explanation of our results from M-theory.
In section 6 we discuss the connection with 6d superconformal indices and in particular
compute the superconformal index for coincident M5 branes.  In section 7 we present our conclusions.
Some more technical aspects of the paper are presented in Appendices A,B and C.

\section{SCFT on squashed $ S^3 $ and open topological strings}

One of the common themes that have emerged in the study of superconformal theories in various dimensions is the important role played by the BPS states that arise when one moves away from the superconformal fixed point (see \cite{Iqbal:2012} and references therein). 

In particular it was shown in \cite{Iqbal:2012} that the superconformal index
in diverse dimensions is deeply related to BPS spectrum and this data can be used to fully compute the index in ${\cal N}=2$ theories
in $d=3 $ and ${\cal N}=1$ theories in $d=5$.   These correspond to partition functions
on $S^2\times S^1$ and $S^4\times S^1$ respectively.  Here we are interested in computing
the partition functions of these theories on $S^3$ and $S^5$, respectively.  To this
end, it is instructive to review the case of $ \scN = 2 $ superconformal theories on the squashed three-sphere $ S^3_b $. This class of theories is particularly simple, since away from the superconformal point only a finite number of BPS particles appear, which are in one-to-one correspondence with the electrically charged fields of the SCFT. The full partition function for these theories has been computed exactly \cite{Hama:2011ea, Imamura:2011wg} and indeed we will see that it can be reinterpreted in terms of contributions coming from the BPS particles (as occurs
in a similar context in \cite{Cecotti:2010fi, Cecotti:2011iy, Pasquetti:2011fj}).

We can write the squashed three-sphere geometry in terms of variables $ (z_{1},z_{2})\in \bC^2 $ as
\[ \omega_1^2 |z_1|^2+\omega_2^2 |z_2|^2 = 1.\]
For $ \omega_1\neq \omega_2 $, the $ SO(4) $ isometry group of $ S^3 $ gets broken to $ U(1)\times U(1) $. The ratio of the equivariant parameters for the two rotations is $ \tau = b^2=\omega_1/\omega_2 $.

We now recall the partition function for superconformal gauge theories on the squashed three-sphere, whose gauge and matter content are provided respectively by vector and chiral multiplets. Away from the superconformal point, many of these theories can be constructed from M-theory as the worldvolume theories of M5-branes wrapping $ S_b^3 $ times a Lagrangian submanifold of an appropriately chosen Calabi-Yau threefold $ X $. The geometry of $ X $ determines the BPS content of the theory, and the superconformal theory is recovered in the IR (shrinking the size of the Lagrangian to zero).

Let $ \mathfrak{g} $ be the Lie algebra of the gauge group $ G $, and $ \mathfrak{h} $ its Cartan subalgebra. Let $ h_i,\; i= 1,\dots, \text{ rank}(G) $ be a basis for $ \mathfrak{h} $. We denote a generic element of $ \mathfrak{h} $ by $ \phi = \sum \phi_i h_i $, and for an arbitrary weight $ \nu $ of $ \mathfrak{g} $ we write $ \phi_\nu = \langle \nu,\phi\rangle $. By localization, the computation of the partition function of the SCFT reduces to an integral over $ \mathfrak{h} $, with contributions from one-loop determinants for the chiral and vector multiplets:
\begin{equation} Z_{S_b^3}=\int d\phi \prod_{\beta\in\Delta_+} \phi_\beta^2  \cdot Z_0(\phi)\cdot Z_{vect}^{1-loop}(\phi)\cdot Z_{chiral}^{1-loop}(\phi), \label{eq:s3bpartition}\end{equation}
where $ \Delta_+ $ is the set of positive roots of $ G $. The classical action can contain Chern-Simons and FI terms, and produces a factor of
\begin{equation} Z_0(\phi) = e^{-\frac{\pi i}{2} k_i \phi_i^2 + 2\pi i \xi_i \phi_i },\label{eq:Z0}\end{equation}
where $ k_{i} $ is the CS level and $ \xi_i\in \bR $ is the FI-term.  For abelian factors
we can also have additional off-diagonal CS interactions as well as
mixed CS terms with flavors symmetries.

If we include matter fields in a (not necessarily irreducible) representation $ R $ of the gauge group $G$, for each weight in $ R \,$ we obtain a chiral multiplet. The one-loop contribution to the partition function is
\begin{align} Z_{chiral}^{1-loop}(\phi) &= \prod_{\mu\in R}\prod_{j,k \geq 0} \frac{(j+1/2)\omega_1+(k+1/2)\omega_2+i\phi_\mu}{(j-1/2)\omega_1+(k-1/2)\omega_2-i\phi_\mu}\notag\\
&=\prod_{\mu\in R}S_2^{-1}\left(i\phi_\mu +(\omega_1+\omega_2)/2|\omega_1,\omega_2\right),\end{align}
where the double sine function $ S_2(z |\omega_1,\omega_2) $ is defined in Appendix \ref{sec:multisine}. The vector multiplet, on the other hand, contributes a factor of (taking
into account the shift in spin $s=1/2$)
\begin{align}Z_{vect}^{1-loop}(\phi) &=\prod_{\beta \in \Delta}\frac{1}{i\phi_\beta} \prod_{j,k \geq 0}\frac{ j\omega_1+k\omega_2 +i\phi_\beta}{(j+1)\omega_1+(k+1)\omega_2-i\phi_\beta}\notag\\
&=\prod_{\beta\in\Delta}\frac{1}{i\phi_\beta} S_2(i\phi_\beta +\omega_1+\omega_2|\omega_1,\omega_2),\end{align}
where by $ \Delta $ we mean the set of roots of $ G $.
Note that for a spin $s$ field we get a shift of 
$$({1\over 2}+s,{1\over 2} +s)\cdot (\omega_1,\omega_2).$$
Putting all the pieces together, the partition function is
\[ Z_{S_b^3} = \int d\phi \; e^{-\frac{i\pi k\phi_i \phi_i}{2}+2\pi i \xi_i \phi_i}\prod_{\beta\in\Delta}S_2(i\phi_\beta +\omega_1+\omega_2|\omega_1,\omega_2) \prod_{\mu\in R} S_2^{-1}(i\phi_\mu+(\omega_1+\omega_2)/2 |\omega_1,\omega_2). \]
Thus to each multiplet $ \alpha $ corresponds a factor of $ S_2(z_\alpha|\omega_1,\omega_2)^{\pm1} $, where the argument of the double sine function depends on the data attached to the multiplet.  
Note that for the vector multiplet the $\prod_{\beta \in \Delta}S_2(i\phi_\beta +\omega_1+\omega_2 |\omega_1,\omega_2)$ is equal to
a $q$-deformed Vandermonde.  The double sine has simple modular transformation under the $S$ transformation of $SL(2, \bZ)$.  Indeed, when $ \tau = \omega_1/\omega_2\in \bH$ the double sine function can be written in the following suggestive form \eqref{eq:s2factorized}:
\begin{align} &S_2(z_\alpha +(\omega_1+\omega_2)/2 | \omega_1,\omega_2)\notag\\
&\qquad=\exp\left(\frac{\pi i}{2} B_{2,2}(z_\alpha+(\omega_1+\omega_2)/2|\omega_1,\omega_2)\right)\frac{\prod_{j=0}^\infty (1-e^{\zeta_\alpha+\pi i +2\pi i (j+1/2)\tau)})}{\prod_{j=0}^\infty (1- e^{\hat\zeta_\alpha+\pi i+2\pi i(j+1/2)\hat\tau})}&\notag\\
&\qquad=e^{\frac{\pi i}{2\tau}(\zeta_\alpha/2\pi i)^2-\frac{\pi i}{24}(\tau+1/\tau)}\ \frac{\prod_{j=0}^\infty (1-e^{\zeta_\alpha+\pi i +2\pi i (j+1/2)\tau)})}{\prod_{j=0}^\infty (1- e^{\hat\zeta_\alpha+\pi i+2\pi i(j+1/2)\hat\tau})},\notag\\
&\qquad=e^{\frac{\pi i}{2\tau}(\zeta_\alpha/2\pi i)^2-\frac{\pi i}{24}(\tau+1/\tau)}\ \frac{\prod_{j=0}^\infty (1+e^{\zeta_\alpha}q^{j+1/2})}{\prod_{j=0}^\infty (1+ e^{\hat\zeta_\alpha}{\hat q}^{j+1/2})},\label{eq:doublesinemodular}\end{align}
where we have defined $ \zeta_\alpha = 2\pi i z_\alpha/\omega_2 $, $\; \hat \zeta_{\alpha} = \zeta_\alpha/\tau $, $ \;\hat \tau = -1/\tau $, and $q={\rm exp}(2\pi i\tau)$ and ${\hat q}={\rm exp}(-2\pi i/\tau)$.
The exponential prefactors come from the $ (2,2) $ multiple Bernoulli polynomial \eqref{eq:Bernoulli22},
\[ B_{2,2}(z_\alpha|\omega_1,\omega_2) = \frac{z_\alpha^2}{\omega_1\omega_2} - \frac{\omega_1+\omega_2}{\omega_1\omega_2}z_\alpha+\frac{\omega_1^2+\omega_2^2+3\omega_1\omega_2}{6\omega_1\omega_2}.\]
Under an S modular transformation that takes $ \tau\to\hat\tau $ and $ \zeta_\alpha\to \hat \zeta_\alpha $,
\[ S_2(z_\alpha +(\omega_1+\omega_2)/2 | \omega_1,\omega_2)\to S_2(z_\alpha +(\omega_1+\omega_2)/2 | \omega_1,\omega_2)^{-1}.\]
On the other hand, the double sine function does not transform into itself under the T transformation $ \tau \to \tau+1 $, so we cannot complete this to a full $ SL(2,\bZ) $ action.

We would now like to clarify the relation with BPS states and open topological string theory. For this purpose, it is convenient to strip away the prefactors from the double sine function and define
\begin{equation} \scS_2(z|\omega_1,\omega_2) = \exp\left(-\frac{\pi i}{2}B_{2,2}(z|\omega_1,\omega_2)\right)S_2(z|\omega_1,\omega_2).\label{eq:s2stripped}\end{equation}
Using the building block of the double sine function we can write down the contribution of
particles of charges $n_i,n_j$ under $U(1)$ gauge factors and flavor factors respectively with
central terms $(x_i,m_j)$ (before gauging) and spins $s$:
$$\scS_2\big((n_ix_i+n_jm_j)+({1\over 2} +s)(\omega_1+\omega_2)|\omega_1,\omega_2\big)^{-(-1)^{2s}}$$
Thus we would get for many particles a partition function of the form:
$$Z=e^{Q(x_i,m_j)}\cdot \prod_a \scS_2\big((n^a_ix_i+n^a_jm_j)+({1\over 2} +s_a)(\omega_1+\omega_2)|\omega_1,\omega_2\big)^{-(-1)^{2s_a}}$$
where we have included the prefactor (involving the exponential of the quadratic form) which
is added at the end depending on the FI terms and the CS levels (see \cite{Closset:2012vp} for
a thorough discussion of these terms). 
To obtain the final partition function we have to integrate over the scalars in the $U(1)$
vector multiplets leading to
$$Z_{S^3}=\int dx_i Z(x_i,m_j,\tau).$$
In the next section we discuss how this can be presented in the context of 3d
theories living on M5 branes wrapped on special Lagrangian 3-cycles, using
open topological string amplitudes.

\subsection{Topological String Reformulation}

We now use topological strings to reformulate this partition function (see also
\cite{Pasquetti:2011fj}).
It is known that open topological strings captures the BPS content of M5 branes
wrapped on special Lagrangian cycles of Calabi-Yau threefold \cite{Ooguri:1999bv}.  For simplicity
we will focus on the unrefined case here (but will extend the discussion to the
refined case when considering the closed string sector).  Consider M-theory
compactification on a Calabi-Yau threefold, and consider a number of M5 branes
wrapping some special Lagrangian cycles.  Then M2 branes ending on M5 branes constitute
the BPS states of the theory.
The partition function of topological strings captures this.  In particular we have (up to
quadratic exponential prefactor)\footnote{We are always free to rescale the arguments of the double sine function  $ z, \omega_1,\omega_2 $ by a common factor. When comparing to topological strings, we choose a gauge where $ \omega_2 = 1 $.}:
$$Z_{top}^{open}=\prod_a\prod_{k=0}^{\infty} (1-q^{k+s_a+{1\over 2}} e^{2\pi in_i^ax_i+2\pi i n_j^a m_j})^{N_{n_i,n_j,s_a}(-1)^{2s_a+1}}$$
For our purposes it is more convenient to define a slightly shifted version of the topological
string amplitude given by
$${\tilde Z}_{top}^{open} =\prod_a\prod_{k=0}^{\infty} (1-(-1)^{2s_a+1}q^{k+s_a+{1\over 2}} e^{2\pi i n_i^ax_i+2\pi i n_j^a m_j})^{N_{n_i,n_j,s_a}(-1)^{2s_a+1}}$$
$$=Z_{top}^{open}(\tau+1),$$
where $q={\rm exp}(2\pi i \tau)$ and $N_{n_i,n_j,s_a}$ denote the number of
BPS states with the corresponding charges as spin. 
We will drop the tilde in the rest of the paper as we will be mainly discussing this shifted version.
The unshifted version can be recovered by shifting the $\tau$ back.  

We now simply ask what would the partition function of this
theory be if we were to put it on the squashed $S^3$?   Even though we have
no a priori Lagrangian description of this theory we will assume, as in \cite{Iqbal:2012},
that the BPS states can
be treated as elementary degrees of freedom.  Using the fact that double
sine computes the corresponding term we would thus naturally get
$$Z=e^{Q(x_i,m_i)} \cdot \prod_{n_i,n_j,s_a} \scS_2\big((n^a_ix_i+n^a_jm_j)+({1\over 2} +s_a)(\omega_1+\omega_2)|\omega_1,\omega_2\big)^{N_{n_i,n_j,s}(-1)^{2s_a+1}},$$
where we have included the prefactor involving the quadratic classical term $Q$ of the topological string.  Using
the product representation of the double sine function and the form of $Z_{top}^{open}$
we can rewrite this entirely in terms of the topological string partition function as
$$Z^{open}_{np}={Z_{top}^{open}(x_i,m_j;\tau)\over Z_{top}^{open}(x_i/\tau,m_j/ \tau;-1/ \tau )}$$
and we can view this as a non-perturbative definition of topological string. Then
 the partition function on squashed $S^3$ is given by
$$Z_{S^3}=\int dx \;e^{Q(x_i,m_i)}\cdot Z^{open}_{np}=\int dx_i \;e^{Q(x_i,m_i)}\cdot {Z_{top}^{open}(x_i,m_j;\tau)\over Z_{top}^{open}(x_i/\tau,m_j/\tau;-1/\tau )}$$
where by definition what we mean by $Z_{top}$ at $-1/\tau$ is the product expression
we have given.  Notice that the factor of $(-1)^{s}$ in the expansion, which for even
$s$ does not seem to affect the perturbative $Z_{top}$, will be relevant
under the $\tau \rightarrow -1/\tau$, which we include in the definition of $Z_{top}$
at $-1/\tau$.

As we have seen, when $ \text{Im } \tau > 0 $,
\[Z^{open}_{np}(\dots;\tau)=Z_{top}^{open}(\dots|\tau)/Z_{top}^{open}(\dots;-1/ \tau);\]
similarly for $ \text{Im } \tau < 0 $,
\[Z^{open}_{np}(\dots;\tau)=Z_{top}^{open}(\dots|1/\tau)/Z_{top}^{open}(\dots;-\tau).\]
But in fact the proposed non-perturbative completion of the open topological string is also valid for $ \tau \in \bR_+ $, i.e. at $ |q| = 1 $, even though the perturbative topological string is ill-defined there. 

\section{Five dimensional superconformal theories}

We saw in the last section that knowing the properties of BPS states of the theory on the squashed three-sphere away from the superconformal fixed point is sufficient to compute the partition function of the SCFT. We now shift our focus to superconformal theories on $ S^5 $ which can be obtained from the compactification of M-theory on a Calabi-Yau threefold. Assuming that in this case too the BPS states account for all the degrees of freedom of the SCFT, we can introduce squashing parameters for $ S^5 $ and propose an exact answer for the partition function (equation \eqref{eq:s5partition}), which includes all gauge theory instanton contributions.  To this
end all we need to know is the contribution of each individual BPS particle to the partition
function and take the product of them over all BPS states, as if they are non-interacting fundamental
degrees of freedom.  Thus the main thing we need to do is to do a computation of the
partition function on squashed $S^5$ for a single BPS particle.

Such a computation has been carried out in \cite{Kallen:2012cs, Kallen:2012va} for certain BPS particles
which appear as the perturbative part of the partition function of $ \scN=1 $ superconformal field theory on $ S^5 $ with non-abelian gauge group and matter in an arbitrary representation $ R $.
We review this result and propose a generalization of it to particles of arbitrary spin.  This is also important
to us for another reason:  As in the 3d case, even if the gauge theory is non-abelian, the computations
can be entirely recast in terms of an integral over the abelian Coulomb branch parameters, where the non-abelian aspects are reflected by the existence of additional BPS states in the computation.
This allows us to formulate our final result in term of an integral over the Coulomb branch.

 In the perturbative computation, the path integral localizes on the Cartan subgroup of the gauge group, and the hyper and vector multiplets, which correspond respectively to the matter and gauge content of the theory, contribute the following one-loop determinants evaluated on the localization locus:
\[ Z_{hyper}^{1-loop}(\phi) = \prod_{\mu \in R}\prod_t\left(t-i\phi_\mu+3/2\right)^{-t^2/2-3/2t-1},\]
where $ \mu $ are the weights in the representation and $ \phi $ is an element of the Cartan, and
\[ Z^{1-loop}_{vect}(\phi) =  \prod_{\beta\in\Delta_+}\prod_{t\neq 0}[(t+i\phi_\beta)(t-i\phi_\beta)]^{t^2/2+3t/2+1}, \]
where $ \Delta_+ $ denotes the positive roots of the gauge group.

In appendix \ref{sec:S31loop} we show that these expressions can be recast in terms of triple sine functions \cite{Jimbo:1996ss, Kurokawa:2003, Narukawa:2003, Nishizawa:2001} as
\begin{equation} Z_{hyper}^{1-loop}(\phi) = \prod_{\mu \in R}S_3^{-1}(i\phi_\mu + 3/2  | 1, 1, 1)\label{eq:zhyper}\end{equation}
and
\begin{equation}Z_{vect}^{1-loop}(\phi) =  \prod_{\beta \in \Delta_+} (i\phi_\beta)^{-2}\prod_{\beta \in \Delta_+} S_3(i\phi_\beta  | 1, 1, 1)S_3(i\phi_\beta + 3 | 1, 1, 1),\label{eq:zvect}\end{equation}
up to a prefactor which can be reabsorbed into the cubic prepotential. The triple sine function is defined as a regularized infinite product over three indices:
\[ S_3(z | \omega_1,\omega_2,\omega_3) \sim \prod_{n_1,n_2,n_3 = 0}^\infty (n_1\omega_1+n_2\omega_2+n_3\omega_3+z)((n_1+1)\omega_1+(n_2+1)\omega_2+(n_3+1)\omega_3 - z)\]
(the precise definition and several important properties of this function are collected in Appendix \ref{sec:multisine}). From this expression it is clear that the one-loop determinants for the theory on $ S^5 $ are evaluated at a very degenerate choice of parameters for the triple sine. In the theory on $ S^3 $ an interesting deformation was obtained by introducing squashing parameters $ \omega_{1,2} $, and the one-loop determinants were found to be built out of factors of $ S_2(z | \omega_1,\omega_2) $. In our current setup, it is also very natural to move away from this limit and consider an analogous deformation by three parameters $ \omega_{1,2,3} $. That is, we conjecture that one can formulate a deformation of the theory on squashed $ S^5 $, which can be embedded in $ \bC^3 $ as
\[ \omega_1^2 |z_1|^2+\omega_2^2 |z_2|^2+\omega_3^2 |z_3|^2 = 1,\]
and that each occurrence of $ S_3(z | 1, 1, 1) $ gets replaced by $ S_3(z | \omega_1, \omega_2,\omega_3) $. The $ SO(6) $ isometry of $ S^5 $ gets broken to $ U(1)^{(1)}\times U(1)^{(2)}\times U(1)^{(3)} $, where $ U(1)^{(i)} $ corresponds to rotation of the $ z_i $-plane. The ratio of the equivariant parameters for $ U(1)^{(i)} $ and $ U(1)^{(j)} $ is given by $ \omega_i/\omega_j $.

The hyper and vector multiplet one-loop determinants become
\[ \scZ_{hyper}^{1-loop}(\phi) = \prod_{\mu \in R}S_3^{-1}(i\phi_\mu + \omega_1/2+\omega_2/2+\omega_3/2  | \omega_1, \omega_2, \omega_3),\]
and
\[   \scZ_{vect}^{1-loop}(\phi) = \prod_{\beta \in \Delta_+} (i\phi_\beta)^{-2}\prod_{\beta \in \Delta_+} S_3(i\phi_\beta  | \omega_1, \omega_2, \omega_3)S_3(i\phi_\beta + \omega_1+\omega_2+\omega_3 | \omega_1, \omega_2, \omega_3).\]
Putting all these contributions together, the perturbative
contribution to the partition function (choosing units where the radius of $S^5=1$) is
\begin{align} Z_{S^5}^{pert} &= \int_{Cartan} d\phi \; \big(\!\!\prod_{\beta\in\Delta_+} \phi_\beta^2\big) \scZ_0(\phi) \scZ_{hyper}^{1-loop}(\phi)\scZ_{vect}^{1-loop}(\phi)\\
&=\int_{Cartan} \scZ_0(\phi) \prod_{\beta\in\Delta_+} S_3(i\phi_\beta  | \omega_1, \omega_2, \omega_3)S_3(i\phi_\beta + \omega_1+\omega_2+\omega_3 | \omega_1, \omega_2, \omega_3)\ \cdot\notag\\
&\hspace{1.03in}\quad\prod_{\mu \in R}S_3^{-1}(i\phi_\mu + \omega_1/2+\omega_2/2+\omega_3/2  | \omega_1, \omega_2, \omega_3),\notag\end{align}
where
$$ \scZ_0(\phi) = \exp\left[{1\over \omega_1\omega_2\omega_3 }\left(i\frac{4\pi^3}{g_{YM}^2} \text{Tr}\phi^2+{ik\over 24\pi^2}\text{Tr}\phi^3 \right)\right],$$
which comes from the tree level Lagrangian (where we have included the effect of $\omega_i$ being turned on).  Notice that this term is the exponential of a cubic polynomial
$Z_0={\rm exp}[C(\phi,{1/g_{YM}^2})]$ where $C$ captures the cubic content
of the prepotential term, where we view $1/g_{YM}^2$ as a scalar in an ungauged vector multiplet.

Just as in the 3d case the non-abelian measure factors have disappeared and  we can interpret the integrand as the contribution of the electric BPS states in an abelian theory,
as we go away from the conformal fixed point on the Coulomb branch.   However, unlike in the 3d case, here there are more BPS states than those captured by the perturbative content of the theory.
In fact, the five-dimensional theory will have an infinite number of BPS states, 
including ones which carry instanton charge.
 Our proposal is that the full partition function on squashed $ S^5 $ is simply given by
 the contribution over {\it all} BPS states and not just the electric ones.  In other words, we propose:
\begin{equation}Z_{S^5} = \int d\phi \;\scZ_0(\phi) \prod_{\alpha\in BPS} \scZ_{\alpha}(z_\alpha|\omega_1,\omega_2,\omega_3),\label{eq:s5partition} \end{equation}
where each $ \scZ_{\alpha} $ is a contribution from a BPS particle written in terms of triple sine function
(and its generalization), and 
$\scZ_0(\phi) = e^{C(\phi,m)}$
is the effective semi-classical contribution and is a polynomial of degree $3$ in $\phi $ and  $ m$.   By $\scZ_{\alpha}$ we mean the determinant contributions coming from the
individual BPS states with the exponential  prefactor stripped off (see the next section for more details).
This proposal fits naturally with the computation in \cite{Kallen:2012va, Kallen:2012cs,Kim:2012av}
where the main missing ingredient was the contribution of instantons to the partition function.
Here we are proposing that the BPS content of the theory, which includes instanton charged states,
completes the computation.

In the case where the superconformal theory comes from a Calabi-Yau threefold, $C$ can be related
to the classical properties of the CY and captures the classical prepotential term, as well as genus 1 corrections
which are linear in $\phi $ and  $ m$.  In the unrefined case $C$ is simply given by
$$C(\phi,m)={1\over 6 \lambda^2}\int_{CY}J\wedge J\wedge J +{1\over 24}({1\over \lambda ^2}-1)\int_{CY} J\wedge c_2, $$
where $J(\phi,m)$ denotes the Kahler form on the CY which is parameterized by $\phi,m$ and $c_2$ is the second
Chern class of the CY where the genus 0 piece can be read off from \cite{Candelas:1990rm,Hosono:1994ax} 
 and the genus 1 piece from \cite{Bershadsky:1993ta}.  In the refined case where $\tau_1+\tau_2\not=0$ this becomes\footnote{We have used the unrefined case together with $SL(3,{\bf Z})$ invariance of the classical prepotential, up to sign, to predict this structure.  One should be able to derive this
 directly from the definition of the refined topological string \cite{Dijkgraaf:2006um}.}
$$C(\phi,m)={1\over 6 \tau_1\tau_2}\int_{CY}J\wedge J\wedge J -{1\over 24 } ({\tau_1\over \tau_2}+{\tau_2\over \tau_1} +{1\over \tau_1\tau_2}+3) \int_{CY} J\wedge c_2.$$
We will choose normalizations where the Kahler class is given by $2\pi iT$.  In this
normalization we can write this as
$$C(T)={-2\pi i}\bigg ({CT^3\over 6\tau_1\tau_2}-{c_2\cdot T\over 24}({\tau_1\over \tau_2}+{\tau_2\over \tau_1} +{1\over \tau_1\tau_2}+3)\bigg)$$
where 
$$CT^3=C_{ijk}T^iT^jT^k,\qquad  c_2\cdot T=c_2^i T^i,$$
 and $C_{ijk}$ denotes the triple intersection
and $c_2^i$ the second Chern class in this basis.
   In the next section we show how topological strings capture this partition function
elegantly, leading on the one hand to the full partition function for ${\cal N}=(1,0)$ theories obtained
by compactification of M-theory on toric CY threefold, and on the other hand to a non-perturbative
definition of topological string.

\section{Non-perturbative topological strings and the partition function on $S^5$}
\label{sec:nonperturbative}

Consider M-theory on Calabi-Yau threefolds.  It is known that topological strings capture the BPS content of  M2 branes wrapped over 2-cycles of the Calabi-Yau \cite{Gopakumar:1998ii, Gopakumar:1998jq}.  Furthermore, in the case
of toric threefolds (which lead to ${\cal N}=(1,0)$ theories of interest to us here) we can consider
a refinement of the BPS counting \cite{Hollowood:2003cv}.  The relation between the topological
string partition function and BPS state counting is given by
$$Z^{top}=\prod_{s_1,s_2,k_i,l_j}\prod_{m,n=0}^{\infty} (1-q^{m+s_1+{1\over 2}}t^{n-s_2+{1\over 2} }{\rm e}^{2\pi i( t_i k_i+m_j l_j)})^{(-1)^{2s_1} N_{s_1,s_2,k_i,l_j}}$$

Note that we have stripped off the classical terms, and below
when we restore the classical pieces we will make it clear. 
Here $q={\rm exp}(2\pi i \tau_1), t={\rm exp}(-2\pi i\tau_2)$ are the coupling constants of the refined topological string,  
 the $N_{s_1,s_2,k_i,l_j}$ are the BPS degeneracies, where $(k_i,l_j)$ denotes the
 gauge and flavor charges of the BPS states and is an element of $H_2$ of the CY where the M2 brane wraps to give rise to BPS state.  Here
$k_i$ corresponds to charges of normalizable Kahler classes $t_i$, and flavor charge $l_j$ corresponds
to non-normalizable Kahler classes $m_j$. The $s_i$ give the $(s_1,s_2)=(J_{12},J_{34})$ content of the
$SO(4)$ rotation group in 5 dimensions.    Namely, viewing $SO(4)=SU(2)_L\times SU(2)_R$
each BPS state is given by
$$I_L\otimes (j_l,j_r)$$
where 
$$I_L =[({1\over 2},0)+2(0,0)]$$
and the $s_i$ just capture the spin content (not including the $I_L$ factor):
$$-j_l\leq {s_1-s_2\over 2} \leq j_l,\qquad -j_r\leq {s_1+s_2\over 2}\leq j_r$$
It will be useful for us to slightly change the definition of topological strings (as in the open sector
discussed in the 3d context) by shifting\footnote{We can shift either $\tau_1$ or $\tau_2$ since
$2s_1=2s_2$ mod 1.  Note that this shift is equivalent to insertion of $(-1)^F$ and will
be explained in section 5.} one of the couplings by 1:
$${\tilde Z}^{top}=\prod_{s_1,s_2,k_i,l_j}\prod_{m,n=0}^{\infty} (1-(-1)^{2s_1+1}q^{m+s_1+{1\over 2}}t^{n-s_2+{1\over 2} }{\rm e}^{2\pi i(t_i k_i+m_j l_j)})^{(-1)^{2s_1} N_{s_1,s_2,k_i,l_j}}$$
$$=Z^{top}(t_i,m_j; \tau_1+1,\tau_2)$$
Since we will be mainly dealing with this object we will be calling
it $Z^{top}$ and drop the tilde.  Of course one can recover the usual
definition of topological string by shifting back the coupling by 1.

In order to connect this to the partition function on $S^5$ we need to know how each
field contributes to the partition function.  Consider a field with spins $(s_1,s_2)$ (coming
as part of a BPS multiplet).  Then we already know that when $(s_1,s_2)=0$ the contribution
is given by a shifted triple sine function:
$$S_3^{-1}\bigg (z+({1\over 2},{1\over 2},{1\over 2})\cdot (\omega_1,\omega_2,\omega_3)\big|\omega_1,\omega_2,\omega_3\bigg)$$
Moreover for a vector multiplet $(0,1/2)$ which has $(s_1,s_2)=(\pm {1\over 2},\pm {1\over 2})$ we get
$$S_3\bigg(z+[({1\over 2},{1\over 2},{1\over 2})\pm ({1\over 2},{1\over 2},{1\over 2})]\cdot (\omega_1,\omega_2,\omega_3)\big|\omega_1,\omega_2,\omega_3\bigg)$$

Now comes the main point. The connection to non-perturbative topological strings come to life thanks to a remarkable formula (equation \eqref{eq:s3factorized}) for the triple sine function:
\begin{align} &\exp\left(-\frac{\pi i}{6}B_{3,3}(z+\Delta |\omega_1,\omega_2,\omega_3)\right)S_3^{-1}(z+\Delta  | \omega_1,\omega_2,\omega_3)\nonumber\\
&\quad=\frac{\prod_{j,k=0}^\infty (1+e^{2\pi i T+2\pi i(j+1/2)\tau_1-2\pi i (k+1/2)\tau_2})}{\prod_{j,k=0}^\infty (1+e^{2\pi i \hat{T}+2\pi i(j+1/2)\hat{\tau}_1-2\pi i (k+1/2)\hat{\tau}_2})\cdot\prod_{j,k=0}^\infty (1+e^{2\pi i\tilde{T}+2\pi i(j+1/2)\tilde\tau_1-2\pi i (k+1/2)\tilde\tau_2})}\nonumber\\
&\quad=\frac{\prod_{j,k=0}^\infty (1+e^{2\pi i T}q^{j+1/2}t^{ k+1/2})}{\prod_{j,k=0}^\infty 
(1+e^{2\pi i \hat T}{\hat q}^{j+1/2}{\hat t}^{ k+1/2})\cdot\prod_{j,k=0}^\infty (1+e^{2\pi i \tilde T}{\tilde q}^{j+1/2}{\tilde t}^{ k+1/2})},\label{eq:tripleprod}
\end{align}
where we have shifted the argument of the triple sine by the universal term $\Delta= (\omega_1+\omega_2+\omega_3)/2 $, and we set $ T = z/\omega_3 $, $ \tau_1 = \omega_1/\omega_3 $, $ \tau_2 = \omega_2/\omega_3 $, and  also
\begin{align*} (\hat{T},\hat{\tau}_1,\hat{\tau}_2) = (T/\tau_1,-1/\tau_1,\tau_2/\tau_1),\\
(\tilde{T},\tilde{\tau}_1,\tilde{\tau}_2) = (T/\tau_2,\tau_1/\tau_2,-1/\tau_2).
\end{align*}
Furthermore, $q={\rm exp}(2\pi i\tau_1)$ and $t={\rm exp}(-2\pi i\tau_2)$ and similarly for the other variables.
Each infinite product in this expression is convergent when $ \text{Im }\tau_1 > 0 > \text{Im }\tau_2  $, but similar convergent expressions can be obtained in other regions (see Appendix \ref{sec:multisine}). The expression for the triple sine function also includes an exponential prefactor which comes from the (3,3) multiple Bernoulli polynomial \eqref{eq:Bernoulli33} with shifted argument,
\begin{align*} -\frac{\pi i}{6} B_{3,3}(z+(\omega_1+\omega_2+\omega_3)/2|\omega_1,\omega_2,\omega_3)&={1\over \omega_1\omega_2\omega_3}\left[-\frac{\pi i}{6}z^3+\frac{\pi i}{24}(\omega_1^2+\omega_2^2+\omega_3^2) z \right]\\\
&\quad\qquad=-i\pi\left[\frac{T^3}{6}\frac{1}{\tau_1\tau_2}-\frac{T}{24}\frac{1+\tau_1^2+\tau_2^2}{\tau_1\tau_2}\right].
\end{align*}
Taking $ z = z_0= k_i t_i + l_j m_j $ for the hypermultiplets and and $ z = z_0 \pm (\omega_1+\omega_2+\omega_3)/2 $ for the vector multiplets and choosing the gauge $ \omega_3 = 1 $, one finds that the numerator in  \eqref{eq:tripleprod} gives precisely the contributions of the hyper and vector multiplets to the topological string partition function! Similarly when $ s_1=s_2 = s $ and $ z = z_0+s(\omega_1+\omega_2+\omega_3) $ the right hand side of \eqref{eq:tripleprod} becomes
\begin{align*}\frac{\prod_{j,k=0}^\infty (1-(-1)^{2s+1}e^{2\pi iz_0}q^{j+s+1/2}t^{k-s+1/2})}{\prod_{j,k=0}^\infty 
(1-(-1)^{2s+1}e^{2\pi i z_0/\tau_1}{\hat q}^{j-s+1/2}{\hat t}^{k-s+1/2})\cdot\prod_{j,k=0}^\infty (1-(-1)^{2s+1}e^{2\pi iz_0/\tau_2}{\tilde q}^{j+s+1/2}{\tilde t}^{k+s+1/2})}.
\end{align*}
The numerator in this expression also captures the contribution to the topological string partition function of a BPS states with spin $(s, s)$.  It is thus natural to propose that the triple sine function
also gives the determinant for spin $(s,s)$ states.  

This triple product structure involving topological string contributions has a simple generalization for arbitrary spin $(s_1,s_2)$:
\begin{align*}C_{s_1,s_2}(z_0|\tau_1,\tau_2)^{-1}\equiv\hspace{5in}\end{align*}
\begin{align*}
\frac{\prod_{j,k=0}^\infty (1-(-1)^{2s_1+1}e^{2\pi iz_0}q^{j+s_1+1/2}t^{k-s_2+1/2})}{\prod_{j,k=0}^\infty 
(1-(-1)^{2s_1+1}e^{2\pi i z_0/\tau_1}{\hat q}^{j-s_1+1/2}{\hat t}^{k-s_2+1/2})\cdot\prod_{j,k=0}^\infty (1-(-1)^{2s_1+1}e^{2\pi iz_0/\tau_2}{\tilde q}^{j+s_1+1/2}{\tilde t}^{k+s_2+1/2})}.
\end{align*}
which we propose to be giving the determinant contribution for spin $(s_1,s_2)$ states.
Note that for $s_1\not=s_2$ this differs from the triple sine function.
Taking the product over all the BPS states, which we need to do according to our proposal for the computation of the partition function over $S^5$, we obtain
\[ Z(t_i,m_j;\tau_1,\tau_2) = Z_0\cdot \prod_{s_1,s_2,k_i,l_j} C_{s_1,s_2}(z_0|\tau_1,\tau_2)^{(-1)^{2s_1+1}N_{s_1,s_2,k_i,l_j}},\]
 where in the above, in addition to the product over the BPS states, we have included
 the cubic prefactor $Z_0={\rm exp}(C(t_i,m_j; \tau_1,\tau_2))$.  We can rewrite this expression as follows:
\begin{equation} Z(t_i,m_j;\tau_1,\tau_2) = Z_0\cdot\frac{Z_3(t_i,m_j;\tau_1,\tau_2)}{Z_1(t_i,m_j;\tau_1,\tau_2)\cdot Z_2(t_i,m_j;\tau_1,\tau_2)}.\label{eq:Z}\end{equation}
The numerator is precisely the topological string partition function,
\[ Z_3(t_i,m_j;\tau_1,\tau_2) = Z^{top}(t_i,m_j;\tau_1,\tau_2),\]
and we can also relate the two factors in the denominator to the topological string partition function:
\[Z_1(t_i,m_j;\tau_1,\tau_2) = \prod_{s_1,s_2,k_i,l_j}\prod_{j,k=0}^\infty 
(1-(-1)^{2s_1+1}e^{2\pi i z_0/\tau_1}{\hat q}^{j-s_1+1/2}{\hat t}^{k-s_2+1/2})^{(-1)^{2s_1}N_{s_1,s_2,k_i,l_j}}
\]
\[\hspace{-1.35in}=Z'_{top}(t_i/\tau_1,m_j/\tau_2;-1/\tau_1,\tau_2/\tau_1)\]
and
\[Z_2(t_i,m_j;\tau_1,\tau_2) = \prod_{s_1,s_2,k_i,l_j}\prod_{j,k=0}^\infty (1-(-1)^{2s_1+1}e^{2\pi iz_0/\tau_2}{\tilde q}^{j+s_1+1/2}{\tilde t}^{k+s_2+1/2})^{(-1)^{2s_1}N_{s_1,s_2,k_i,l_j}}
\]
\[\hspace{-1.3in}=Z'_{top}(t_i/\tau_2,m_j/\tau_2;\tau_1/\tau_2,-1/\tau_2).\]
The prime signifies that these two factors of the topological string have $ SU(2)_L $ and $ SU(2)_R $ exchanged, which is equivalent to replacing $ (s_1,s_2) $ with $ (-s_1,s_2) $ (or equivalently $ (s_1,-s_2) $) for each BPS state.
In fact, not worrying about regions of convergence, we can use the identity
\[ \prod_{p=0}^\infty (1-Xe^{2\pi i p\gamma}) = \prod_{p=0}^{\infty} (1-Xe^{-2\pi i (p+1)\gamma})^{-1}\]
to rewrite the product of BPS contributions simply as the product of three factors of the topological string partition function:
\[ Z_{top}(t_i,m_j;\tau_1,\tau_2)\cdot Z_{top}(t_i/\tau_1,m_j/\tau_1;1/\tau_1,\tau_2/\tau_1)\cdot Z_{top}(t_i/\tau_2,m_j/\tau_2;\tau_1/\tau_2,1/\tau_2).\]

 Equation \eqref{eq:Z} can be viewed as defining a non-perturbative completion of topological string, in the
 sense that the two additional factors are non-perturbative, as they involve at least
 one $\tau_i \rightarrow -1/\tau_i$.  At the end of this section we will explain the analytic
 properties of $Z$ as a function of $\tau_i$.  Just to complete our discussion,
 in order to compute the $S^5$ partition function we simply have to integrate this
 over the directions in $t_i$:
 $$Z_{S^5}=\int_{t_i}dt_i Z(t_i,m_j;\tau_1,\tau_2).$$
\begin{figure}[t!]
\center
\includegraphics[width=4.5in]{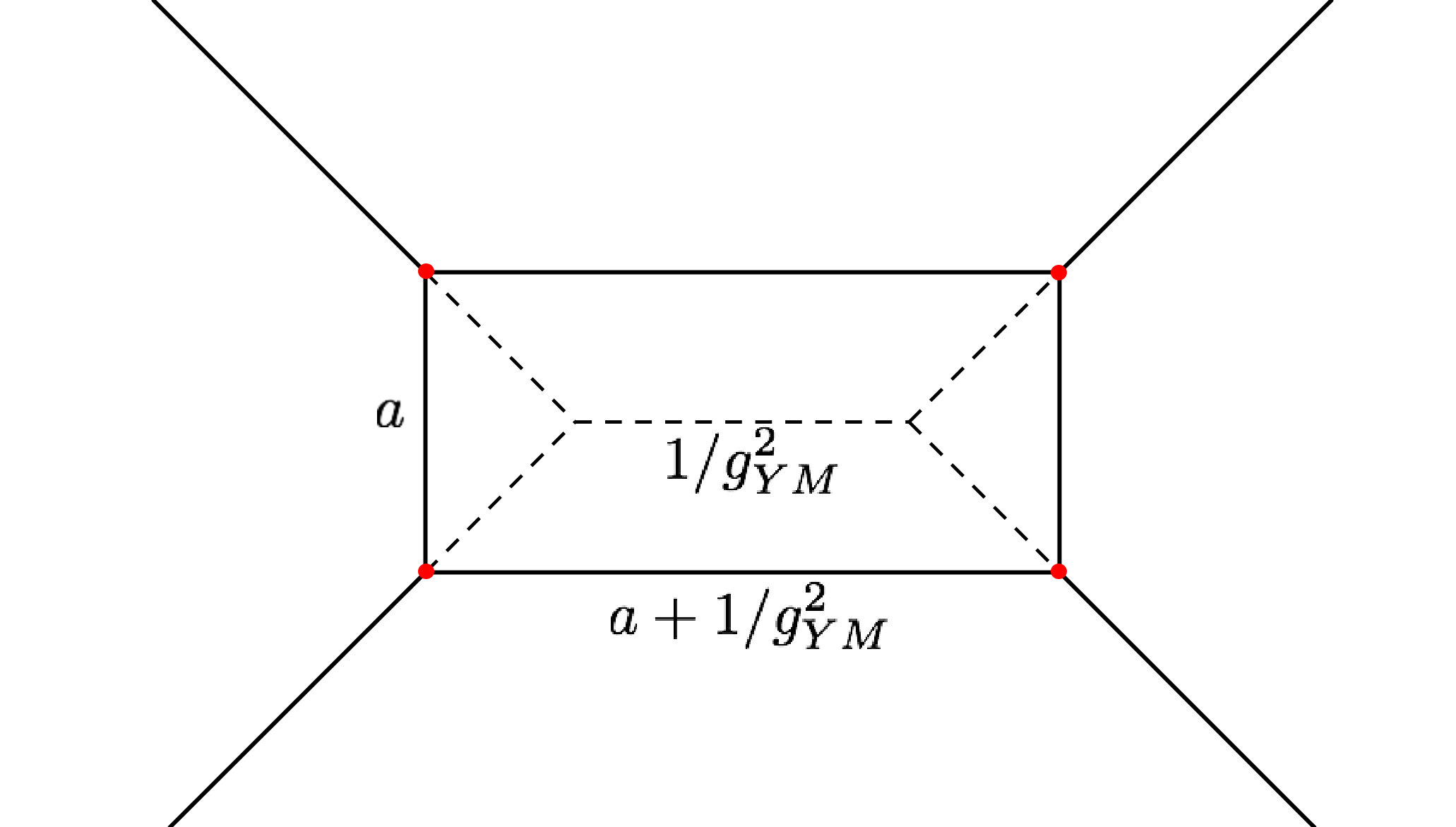}
\caption{$ \bP^1\times \bP^1 $ geometry corresponding to SU(2) theory on the squashed five-sphere. The non-perturbative topological string computed from this geometry is to be integrated over $ a $.}
\label{fig:su2}
\end{figure}
\subsection{Contribution of the massless vector multiplet}
The massless vector multiplets also make a contribution to the partition function.
These contributions do not depend on the moduli but depend on the squashing parameters.
Therefore they can be brought out of the integrals over the Coulomb branch.
These terms are given in the topological string context by powers of the MacMahon function.
If we have $U(1)^r$ gauge theory this leads, as discussed in \cite{Iqbal:2012}, to 
$${(M(q,t)M(t,q))}^{r/2}$$
where
$$M(t,q)=\prod_{i,j=1}^{\infty}(1-q^it^{j-1})^{-1}$$
If we use our prescription to compute the contribution of this factor to the partition function we
get a factor of 
$$\big(S_3(1+\tau_1+\tau_2|1,\tau_1,\tau_2)\cdot S_3(0|1,\tau_1,\tau_2)\big)^{r/2} = S_3(0|1,\tau_1,\tau_2)^r.$$
This has a zero for each $U(1)$ reflecting the fact that we have to delete
the zero mode associated to the Coulomb branch parameters and instead integrate
over it, which is part of the prescription.  This is equivalent to replacing
$S_3$ with its derivative $S_3'$ evaluated at $0$.   In other words, the contributions
for the massless vector multiplet to the partition function is
$$ S'_3(0|1,\tau_1,\tau_2)^{r}\prod_{i=1}^r dT_i.$$
\subsection{An Example:  SU(2) gauge theory}

Here we present one example of how the computation is done.  The case
we focus on is a toric 3-fold that engineers $ SU(2) $ gauge theory coming from the $O(-2,-2)\rightarrow \bP^1\times \bP^1$ geometry.  We consider
the partition function of this theory on the squashed five-sphere.  As discussed, we predict the full partition function to be
\[ Z_{SU(2)} = \int_{a\in i\bR} da\ \frac{e^{C(a,1/g_{YM}^2)}\cdot Z_{SU(2)}(Q_b,Q_f,\tau_1,\tau_2)}{Z'_{SU(2)}(Q_b^{1/\tau_1},Q_f^{1/\tau_1},-1/\tau_1,\tau_2/\tau_1)\cdot Z'_{SU(2)}(Q_b^{1/\tau_2},Q_f^{1/\tau_2},\tau_1/\tau_2,-1/\tau_2)}, \]
where $ Q_f = e^a,\ Q_b = e^{a+ 1/g_{YM}^2} $, and $ Z_{SU(2)}(Q_b,Q_f,\tau_1,\tau_2) $ is the refined topological string partition function for the $ \bP^1\times \bP^1 $ geometry of Figure \ref{fig:su2}, which was obtained in \cite{Iqbal:2007ii} (which is the same as Nekrasov's partition function for the
5d $SU(2)$ theory \cite{Nekrasov:2003rj} with $\epsilon_i=\tau_i$):
\begin{align*}
Z_{SU(2)}(Q_{b},Q_{f},\tau_1,\tau_2):&=\ {[M(q,t)M(t,q)]}^{1/2}\\
&\quad \cdot\sum_{\nu_{1},\nu_{2}}(-Q_{b})^{|\nu_{1}|+|\nu_{2}|}Z_{\nu_{1},\nu_{2}}(t,q,Q_{f})\,f_{\nu_{1},\nu_2}(q,t)
\,Z_{\nu_{2},\nu_{1}}(q,t,Q_{f}), \end{align*}
where
$$q={\rm exp}(2\pi i \tau_1),\qquad t={\rm exp}(-2\pi i \tau_2),$$
\begin{align*}
f_{\nu_{1},\nu_2}(q,t)&=(-1)^{|\nu_{1}|}\Big(\frac{t}{q}\Big)^{\frac{||\nu_{1}^{t}||^{2}-|\nu_{1}|}{2}}\,q^{-\frac{\kappa(\nu_{1})}{2}}\,\,(-1)^{|\nu_{2}|}
\Big(\frac{q}{t}\Big)^{\frac{||\nu_{2}^{t}||^{2}-|\nu_{2}|}{2}}\,t^{-\frac{\kappa(\nu_{2})}{2}},\end{align*}
and
\begin{align*}
Z_{\nu_{1},\nu_{2}}(t,q,Q_{f})&= q^{\frac{||\nu_{1}||^{2}}{2}+\frac{||\nu_{2}^{t}||^{2}}{2}}\widetilde{Z}_{\nu_{1}}(t,q)\widetilde{Z}_{\nu_{2}^{t}}(t,q)
\prod_{i,j}\Big(1-Q_{f}\,t^{i-1-\nu_{2,j}}\,q^{j-\nu_{1,i}}\Big)^{-1},\end{align*}
where
\[ \widetilde{Z}_{\nu}(t,q) =\prod_{s\in
\nu}(1-t^{a(s)+1}q^{\ell(s)})^{-1}\]
and
\[M(t,q)=\prod_{i,j=1}^\infty(1-q^it^{j-1})^{-1}.\]
The classical piece $C(a,{1\over g_{YM}^2})$ is given by
$$C\big(a,{1\over g_{YM}^2}\big)=-{2\pi i\over \tau_1\tau_2}\left({a^{2}\over 2 g_{YM}^2} +{a^3\over 6}\right) +{2\pi i\over 24}\left(-2a+{4\over g_{YM}^2}\right) \left({\tau_1\over \tau_2}+{\tau_2\over \tau_1} +{1\over \tau_1\tau_2}+3\right).$$
The partition function involves sums over Young diagrams. We use the following notation: $ \nu^t $ is the transpose of $ \nu $; $ |\nu| $ denotes the number of boxes in $ \nu $; $ \nu_i $ is the number of boxes in the $ i $-th column of $ \nu $; $ ||\nu||^2 = \sum_i\nu_i^2 $; for a box $ s = (i,j) $ in the $ i $-th column and $ l $-th row of $ \nu $, $ a(s) = \nu_j^t-i $ and $ \ell(s)=\nu_i-j $; and, lastly, $ \kappa(\nu) = 2\sum_{s\in \nu} (j-i)$. Recall that we need to shift $\tau_1 \rightarrow \tau_1+1$ in these formulas to obtain the $Z_{SU(2)}$
appearing in the integrand.

 \subsection{Analytic properties of Z}

The triple sine function (as discussed in Appendix \ref{sec:multisine}) is defined only when all three
$\omega_i$ are in the same half plane.  If this is satisfied, the triple sine function is 
well defined and is an entire function which has zeroes at a lattice of points corresponding
to $n_it_i+k_jm_j=(n_1+{1\over 2})\tau_1+(n_2+{1\over 2})\tau_2+(n_3+{1\over 2})$ (see Appendix \ref{sec:multisine}). Similarly the function $C_{s_1,s_2}(n_it_i+k_jm_j|\tau_1,\tau_2)$ has zeros and poles at values of $ n_it_i+k_jm_j $ which can be read off from equation \eqref{eq:Cprod}.  It is natural
to also expect that $C_{s_1,s_2}$ is well-defined only when all three $\omega_i$ are in the same half plane. 
The non-perturbative topological string partition function is made up of an infinite product of such functions which we conjecture to exist. 

\section{A possible derivation from M-theory}
\label{sec:derivation}

In this section we propose an explanation for the triple product structure that arises when one introduces squashing parameters for $ S^5 $. We start by recalling in more detail the M-theory setup that computes the topological string partition function. We pick a non-compact toric Calabi-Yau threefold $ X $, and take the remaining five-dimensional space to be the Taub-NUT space $ TN $ times the M-theory circle $ S^1 $. We express Taub-NUT space in terms of complex variables $ (z_1,z_2) $ and introduce a twist: as we go around $ S^1 $, we rotate $ (z_1,z_2) \to (e^{2\pi i \tau_1}z_1,e^{2\pi i \tau_2}z_2) $ (and do a compensating twist on $X$ to keep it supersymmetric). We denote 
this twisted space by $ (TN \times S^1)_{\tau_1,\tau_2} $. Then it is known that \cite{Dijkgraaf:2006um}
\[Z_{top}(X,\tau_1,\tau_2) = Z_{M-theory}(X\times TN\times S^1)_{\tau_1,\tau_2}.\] 
The M-theory partition function counts the number of M2-branes wrapping cycles in $ X $, which project to points in Taub-NUT space. When the equivariant parameters are turned on, the particles are concentrated around the origin $ z_1 = z_2 = 0 $.

We can also consider the open string sector of topological strings, which corresponds to adding $ M5 $ branes wrapping a Lagrangian submanifold  $L \subset X$ and the Melvin cigar ($ MC $) subspace of $ (TN \times S^1)_{\tau_1,\tau_2} $ , which has the geometry of  $ S^1\times \bC_{\tau_1}  $. Here $ S^1 $ is the M-theory circle, and $ \bC_{\tau_1} $ is the plane in $ TN $ with rotation parameter $ \tau_1 $ (but we could as well have chosen our M5-branes to fill $ \bC_{\tau_2} $). In topological string theory, wrapping an M5 brane on $ K = MC \times L $ translates to placing a $ \tau_1$-brane on $ L $ \cite{Ooguri:1999bv}  (see \cite{Aganagic:2009cg,Cecotti:2010fi} for a discussion of the refined case).  The problem of counting worldsheet instantons ending on $ L $ translates to counting the states of a gas of M2-branes which wrap two-cycles of $ X $ with boundary on $ L $; the M2 branes project to points on the Melvin cigar. Turning on equivariant parameters again forces these particles to be concentrated at the tip of the cigar, which is located at $ z_1=z_2 = 0 $. Then the M5 brane partition function in this setup is the same as the open topological string partition function:
\[ Z_{M5}(X\times (TN \times S^1)_{\tau_1,\tau_2}, K) = Z^{open}(\vec{t},\vec{x},\tau_2; \tau_1),\]
where $ \vec{t} $ and $ \vec{x} $ denote, respectively, the closed and open string moduli corresponding to $ X $ and $ L $.  In other words, the open topological string theory computes the partition function of the 3d
theory obtained by wrapping an M5 brane on $L$ in the background of $MC$.   Note that for fixed $|z|\not =0$ on ${\bf C}$, the $MC$
has a torus structure, where one circle corresponds to the phase in the $z$-plane and the other
is the circle in the fiber.
Moreover the twisting of the $MC$ as we go around the $S^1$ suggests that changing $\tau$ changes the complex structure of this torus
and it is natural to view this torus as having complex structure $\tau$. 

\begin{figure}[t!]
\center
\includegraphics[width=6in]{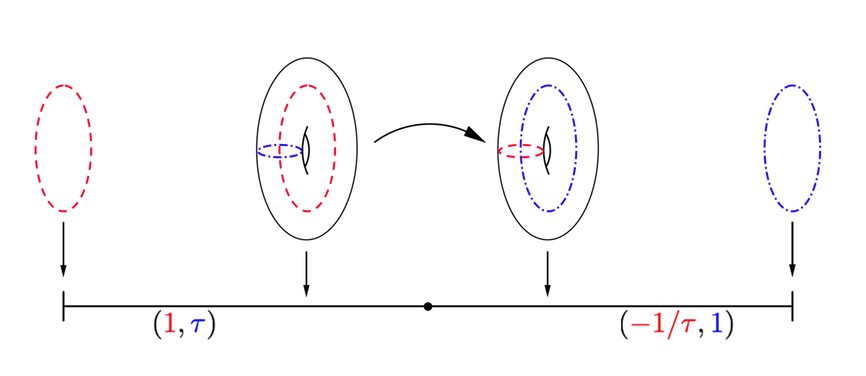}
\caption{Squashed $S^3$ viewed as a torus fibered over the interval. At the ends of the interval one of the two circles degenerates. On the left half of the geometry, as one goes around the red (dashed) circle, the second circle is twisted by $ 2\pi i\tau $. In gluing the left and right halves, one must interchange the two circles of $ T^2 $. On the right half, in going around the blue circle the red circle gets twisted by $ -2\pi i/\tau $.}
\label{fig:fibration}
\end{figure}

To obtain the partition function of the resulting theory on squashed $ S^3 $ we take a second copy of the Melvin cigar, which we denote by  $\widehat{MC}$, and glue it to the first one along the common boundary (as was suggested in the topological string context in \cite{Cecotti:2010fi, Cecotti:2011iy} and discussed in detail in \cite{Pasquetti:2011fj}). This operation can be visualized most clearly by regarding the squashed $ S^3 $ as a torus fibration over the interval, as in Figure \ref{fig:fibration}, and the $T^2$ is the one we have discussed away from the tips of $MC$ and $\widehat{MC}$.
  Each Melvin cigar fills out a solid torus, and we glue the two after performing an $ S $ modular transformation which interchanges the two circles in $\widehat{MC}$. 
  The only subtlety is that we need to ensure that the two cigars are twisted in a compatible way.
In particular the complex structure parameter as seen from the viewpoint of
one tip is different from that of the other end.  This forces us to rescale the rotation parameter for $ \widehat{MC} $ to
\[ \hat \tau_1 = -1/\tau_1.\]
Moreover the topological string has opposite orientation on the $\widehat{MC}$ suggesting
complex conjugation of the topological string amplitude, which is equivalent to inversion of $Z$.
The partition on $ S_b^3 $ then is just the product of the topological string factors on the two hemispheres\footnote{Here we are ignoring the $\tau_2$ dependence which we discuss later in the context of closed strings.},
\[ Z_{S^3} ={ Z_{top}^{open}(\vec{t},\vec{x}, \tau_1)\over Z_{top}^{open}(\vec{t}/\tau_1,\vec{x}/\tau_1, -1/\tau_1)}.\]
The main lesson we extract from the open string case is that for generic choices of the rotation parameters the topological string (or, equivalently, M-theory) computation localizes at the fixed points of the equivariant action on $ \bC^2 $.  In discussing aspects of closed strings we will have to recall that
when we have a more complicated geometry made of patches which look like
$ \bC_{\tau_a,\tau_b}^2\times S^1 $, we would expect by localization to get a contribution of $Z_{top}$ from each patch.
The main new ingredient  is to find the identification of $\tau_1,\tau_2$ between the patches.

\begin{figure}[t!]
\center
\includegraphics[width=4.5in]{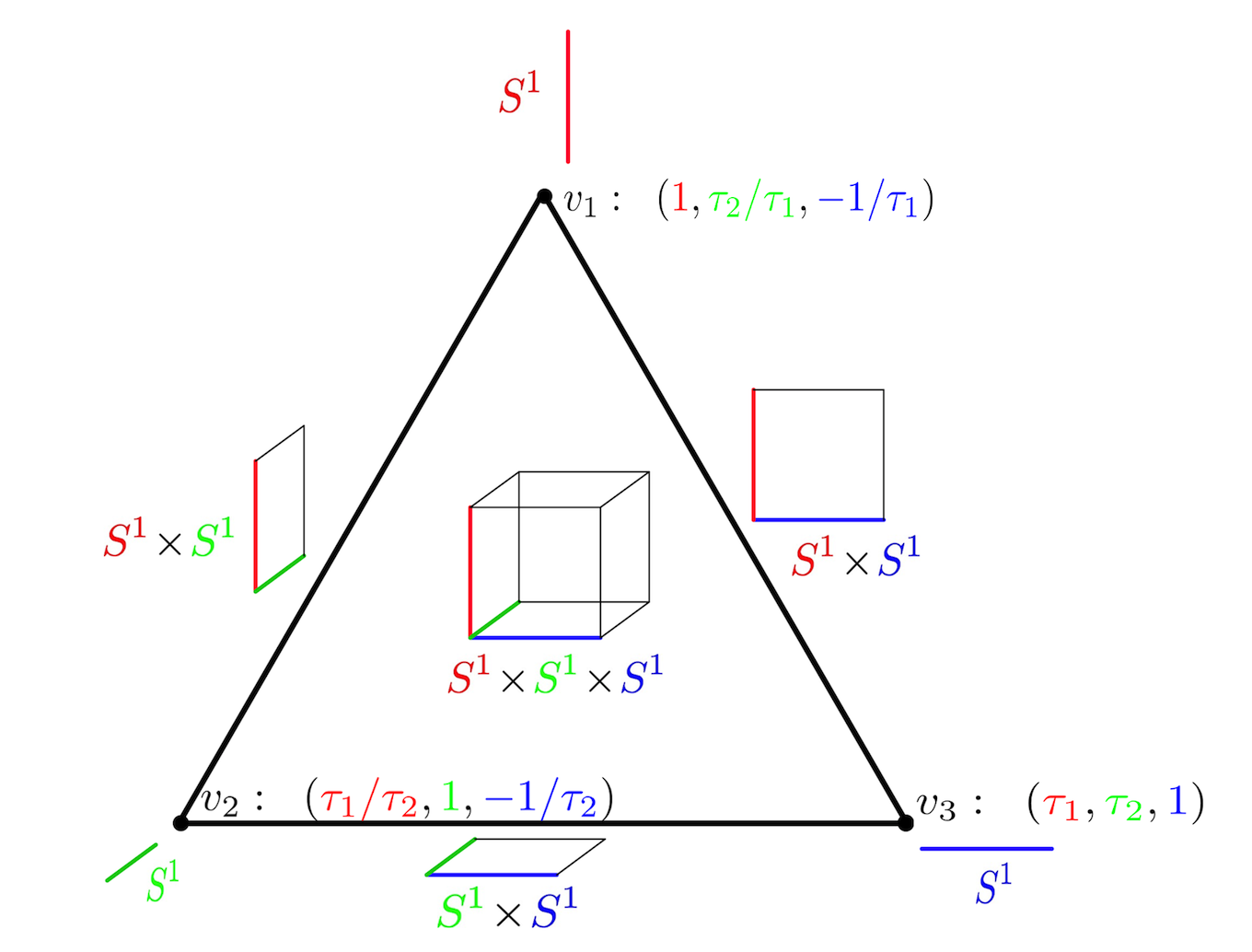}
\caption{Squashed $S^5$ as a $ T^3 $ fibration over a triangle: the cube, whose opposite faces are identified, represents the torus. At the edges of the triangles the torus collapses to a $ T^2 $; at the vertices it collapses to $S^1$. At each vertex we also display the correctly normalized equivariant parameters corresponding to the three circles.}
\label{fig:triangle}
\end{figure}

With this picture in mind, we wish to study the partition function on $S^5$.  We view $S^5$ as a circle
fibration over ${\bf CP}^2$.  Moreover ${\bf CP}^2$ itself can be viewed as consisting of a ${\bf T^2}$
over a triangle, as is familiar in the context of toric geometries (see e.g. \cite{Leung:1997tw}).  Thus we can
think about the squashed five-sphere as a $ S^1\times S^1\times S^1=T^3 $ fibration over a triangle, where each circle in the fiber gets rotated by a different parameter $ \tau_i $ (see Figure \ref{fig:triangle}). In the interior of the triangle all three circles have finite size, but along the edges one of them shrinks to zero size, and the vertices are the points where two of the circles degenerate. We find it convenient to label by $ v_i $ the vertex where the $ i $-th circle of the fiber does not degenerate. We also denote by $ e_{ij} $ the edge that connects $ v_i $ and $ v_j $. It is easy to convince oneself that the neighborhood of $ v_i $ looks like $ S^1_i\times\bC_j\times \bC_k $, where $ i\neq j \neq k $ and each circle in the fiber corresponds to a different factor in the geometry. So from each vertex we expect a contribution of $ Z_{top}^{closed} $\footnote{Up to the factor of $(-1)^F$ because the corresponding $S^1$ in this case
is shrinkable inside $S^5$ and gives a different spin structure compared to the usual case where
$S^1$ is not contractible.
This explains the origin of the shift $\tau_1\rightarrow \tau_1+1$ in the previous
sections.}. To figure out the appropriate parametrization at each vertex, one can start by setting the equivariant parameters to be $ (\tau_1,\tau_2,1) $ at $ v_3 $, so that we get a factor of $ Z_{top}^{closed}(\vec{t},\tau_1,\tau_2) $. We can reach the two other vertices by moving along the edges $ e_{31} $ and $ e_{32} $.  At $ v_1 $ the role of the M-theory circle is played by the first circle, so for the gluing along the edge to be consistent we are required to rescale the equivariant parameters by $ 1/\tau_1 $. This gives us a factor of $ Z_{top}^{closed}(\vec{t}/\tau_1,1/\tau_1,\tau_2/\tau_1) $. Similarly we learn that $ v_2 $ contributes a factor of $ Z_{top}^{closed}(\vec{t}/\tau_2,\tau_1/\tau_2,1/\tau_2) $. Collecting the contributions from the three vertices, we find that M-theory on squashed $ S^5 $ computes
\[ Z_{closed}^{\Delta} =  {Z_{top}^{closed}(\vec{t},\tau_1,\tau_2)\cdot Z_{top}^{closed}(\vec{t}/\tau_1,1/\tau_1,\tau_2/\tau_1)\cdot Z_{top}^{closed}(\vec{t}/\tau_2,\tau_1/\tau_2,1/\tau_2)}.\]

As explained in section \ref{sec:nonperturbative}, we can rewrite this expression in convergent form as
\[ Z_{closed}^{\Delta} =  {Z_{top}^{closed}(\vec{t},\tau_1,\tau_2)\over Z_{top}^{'\,closed}(\vec{t}/\tau_1,-1/\tau_1,\tau_2/\tau_1)\cdot Z_{top}^{'\,closed}(\vec{t}/\tau_2,\tau_1/\tau_2,-1/\tau_2)},\]
where the factors in the denominator are to be computed with the $ SU(2)_L $ and $ SU(2)_R $ spins exchanged.

The non-perturbative open topological string fits very nicely in this picture: the fiber over an edge $ e_{ij} $ consists of two non-degenerate circles $ S^1_i $ and $ S^1_j $, which play inverted roles at the two vertices. This means that over each edge we have a squashed $ S^3 $, so we can get an open sector by wrapping an M5-brane around it and around a Lagrangian submanifold in $ X $. If we do this for the $ e_{13} $ edge we get a contribution of
\[ Z_{open}^{e_{13}} = Z_{open}(\vec{t},\vec{x},\tau_1,\tau_2)/Z_{open}(\vec{t}/\tau_1,\vec{x}/\tau_1,-1/\tau_1,\tau_2/\tau_1).\]
If we were to choose the $ e_{23} $ edge, we would obtain
\[ Z_{open}^{e_{23}} = Z_{open}(\vec{t},\vec{x},\tau_1,\tau_2)/Z_{open}(\vec{t}/\tau_2,\vec{x}/\tau_2,\tau_1/\tau_2,-1/\tau_2).\]

To make this into a rigorous derivation for arbitrary toric Calabi-Yau, we would need to have a way to compactify the full M-theory
on $S^5$, which will necessarily involve some unconventional fields being turned on
(similar to what was found in the 4d case \cite{Dumitrescu:2012ha}).  It is natural to conjecture,
given what we are finding, that such a setup should be consistent, at least in the
case of non-compact Calabi-Yau's.   In the subset of cases where
the CY engineers a gauge theory,
where $Z^{top}$ is identified with the Nekrasov partition function, it should be possible
to rigorously derive this result from the localization arguments in the path-integral.

\section{Superconformal Indices in 6 Dimensions}

It is natural to ask whether the techniques we have introduced can be
used to compute superconformal indices in 6 dimensions.  This
is natural because this involves computations of the amplitudes on
$S^5\times S^1$.  Moreover, compactification on $S^1$ leads to
a 5 dimensional theory, of the type we have studied.  Also, as in the
lower dimensional case studied (such as $S^1\times S^4 $)
turning on the fugacities and supersymmetric rotations of the $S^5$ should
correspond to introducing squashing parameters for $S^5$.

In this section we show how this can be done.  The generic case of interest
is superconformal theories with ${\cal N}=(1,0)$ supersymmetry.  A special
case of these are the $(2,0)$ theories.  We will discuss each one in turn.

\subsection{${\cal N}=(1,0)$ Superconformal Index}

Interacting superconformal theories with ${\cal N}=(1,0)$ supersymmetry are believed to exist.
There are various constructions for them, including small $E_8$ instantons
of heterotic strings \cite{Witten:1995gx}, 5 branes probing ADE singularities \cite{Intriligator:1997dh}
and F-theory constructions on elliptic threefolds with vanishing two-cycles in the base
\cite{Morrison:1996pp, Witten:1996qb}.

The R-symmetry for this case is $Sp(2)$.  Let $R$ denote its Cartan.
The superconformal index in this case can be defined as follows \cite{Bhattacharya:2008zy}:
$$I_{1,0}={\rm Tr}(-1)^F {\bf q}_1^{J_{12}-R}{\bf q}_2^{J_{34}-R}{\bf q}^{J_{56}-R}{\bf M}_i^{F_i}$$
where $J_{ij}$ denote the rotation generators of $SO(6)$ acting on $S^5$, and
$F_i$ are charges associated to flavor symmetries
(where we have only kept the terms which appear non-trivially in the partition function).
The choice of the parameters ${\bf q}_1,{\bf q}_2$ is motivated from connection with
the rotations in 4d, already discussed in the context of 5d theories.

The basic idea, similar to relating the 4d index to 3d partition functions \cite{Imamura,Gadde,Dolan}, is to connect the 6d index to our 5d setup by compactifying this theory on $S^1$.    The only subtlety
is to identify the charges as well as the relation of the parameters in the lower dimensional
theory with the higher dimensional theory.  In the context of compactification of the 6d theory
on a circle, we would need to enumerate
the resulting 5d BPS states  (including winding of 6d BPS strings around the $S^1$) and simply apply the formalism we have developed to this 5d theory.
Here the 5d theory will have a tower of BPS states with a specific structure due to the fact
that it is coming as a KK reduction from a one higher dimensional theory.  If this
theory is dual to M-theory on a CY then from the
perspective of this 5d theory we can enumerate all BPS states using topological strings.
Then using the three
combinations of them and integrating over the scalars in the gauge multiplets yields the
partition function on $S^5$, thus effectively computing the index of the 6d theory.  

Note that from the perspective of the 5d BPS counting, the KK momentum should
appear as a special flavor symmetry.  In the context of F-theory on elliptic CY and its
duality with M-theory upon compactification on $S^1$ (as we will
review below), this will turn out to be the winding number over an elliptic fiber.  We will denote
the Kahler class of the elliptic fiber by $\tau$ and define $q={\rm exp}(2\pi i\tau)$, where
$\tau $ is the Kahler modulus of the elliptic fiber (the reason for this terminology will
become clear later).
Let $M_i={\rm exp}(2\pi im_i)$,
 where $m_i$ denote the non-dynamical fields (coming from non-normalizable
 Kahler moduli).   
 The question is what is the relation between the 5d parameters $q,q_1,q_2,m_i$ with the
 parameters appearing in the 6d index
 ${\bf q},{\bf q}_1,{\bf q}_2,{\bf m}_i$?  A similar situation was studied in the relation between
 superconformal index in 4d and the partition function in 3d \cite{Imamura,Gadde,Dolan}.  In that
 case the squashing parameter are rescaled by a factor of ${\bf R}$, the radius of the circle.
 We propose a similar relation in this case.  Using the fact that the Kahler class of the elliptic
 fiber in F-theory is related to $R$ by
 $$2\pi i \tau = {1\over {\bf R}}$$
 we are led to
 $$({\mathbf \tau},{\mathbf \tau}_1,{\mathbf \tau}_2,{\mathbf m}_j)_{6d}=({-1\over \tau}, \tau_1/\tau,\tau_2/\tau, m_j/\tau)_{5d}$$

In computing the partition function on squashed $S^5$ we need to integrate over
the dynamical fields.
Let $t_i$ denote the scalars associated to the resulting gauge fields
in 5d coming from 6d tensor multiplets, which are normalizable (corresponding
to normalizable Kahler moduli of the CY).  Then we obtain the formula

$$I_{(1,0)}(m_j/\tau ;-1/\tau, \tau_1/\tau ,\tau_2/\tau)=\int dt_i { Z^{top}(t_i,m_j;\tau,\tau_1,\tau_2)\over
Z^{'top}({t_i\over \tau_1},{m_j\over \tau_1}; {\tau\over \tau_1},{-1\over \tau_1},{\tau_2
\over \tau_1})\cdot Z^{' top}({t_i\over \tau_2},{m_j\over \tau_2};{\tau\over \tau_2},{\tau_1\over \tau_2},{-1\over \tau_2})}.$$
This naturally follows from our formalism. It is a general proposal regardless
of whether or not we have a topological string realization of the theory:  The
$Z^{top}$ factor simply denotes the BPS partition function.
 However the question is how to compute
the BPS partition function.  If we can relate it to an actual topological string
then we have techniques for its computation; the most convenient one for this purpose is the F-theory construction, because of the duality between F-theory compactified on $S^1$ and M-theory
on the same space \cite{Vafa:1996xn}.  Thus in 5 dimensions we obtain the theory involving
M-theory on an elliptic 3-fold.  Luckily topological strings on elliptic 3-folds
have very nice properties and have been studied extensively \cite{Klemm:1996hh,Minahan:1997ch, Minahan:1998vr, Hosono:1999qc, Klemm:2012sx,Alim:2012ss}. The relation between 6d and 5d theories via F-theory/M-theory duality has also been studied in  \cite{Bonetti:2011mw}.

As an example, consider the superconformal theory associated with a small $E_8$ instanton.
In the F-theory setup, this corresponds to F-theory with vanishing ${\bf P}^1$ in the base of F-theory
\cite{Morrison:1996pp, Witten:1996qb}.  After compactification on $S^1$, this gives an elliptic 3-fold containing
${1\over 2} K3$ (obtained by the elliptic fibration over the ${\bf P}^1$).  This theory has
10 Kahler classes:  One elliptic fiber class $\tau $, the base $t_b$ and eight mass parameters
$m_i$ (to be identified with the Cartan of $E_8$).  $\tau$ corresponds to momentum and
$t_b$ corresponds to the winding of the 6d tensionless string along
the circle \cite{Klemm:1996hh}.
The unrefined topological
string for this theory was studied in \cite{Klemm:1996hh,Minahan:1997ch,Minahan:1998vr,Hosono:1999qc}.  To obtain
the index for this theory we have to integrate over the $t_b$. 
Similarly a large class of $(1,0)$ theories can be obtained by considering
F-theory where the base contains more blow ups on $\bC^2$ (see \cite{Morrison:2012js}
for a recent discussion related to this).  This would
entail blowing up a multiple of times, each corresponding to a Kahler parameter
$t_i$, which we will have to integrate over in computing the index
(the corresponding $U(1)$ vector multiplet in 5d arises from the 6d tensor multiplet
in the same multiplet as the blow up parameter $t_i$).  A subset of such blowups are the toric
ones.  These are in one-to-one correspondence with 2d Young diagrams \cite{Iqbal:2003ds}.
Elliptic threefolds over these spaces, in the limit of blowing down all the 2-cycles,
should correspond to a $(1,0)$ conformal theory. The case of
a Young diagram with a single row with $k$ entries corresponds to $k$ small $E_8$ instantons.
 It would be interesting to study this large class of $(1,0)$ theories given by a Young diagram.
In particular it should be interesting to compute the corresponding refined topological
strings for this background.   The topological string partition functions for this class of
theories seem to enjoy the following perturbative modular property under the inversion of the
Kahler class of the elliptic fiber \cite{Minahan:1998vr,Hosono:1999qc,Klemm:2012sx,Alim:2012ss}:
$$Z^{top}(t_i,{m_j/ \tau}; -1/\tau,\tau_1/\tau, \tau_2/ \tau)=
Z^{top}(t_i,m_j;\tau,\tau_1,\tau_2).$$
Note the asymmetric role in the modular transformation for the dynamical fields $t_i$ versus
the non-dynamical fields $m_j$  which correspond to
flavor symmetries\footnote{To get this modular transformation, $t_i$ should be suitably
defined by shifting the blow up parameters with a multiple of elliptic fiber \cite{Klemm:2012sx}.}.
In the context of our non-perturbative completion, as we will see later in the context
of the theory of M5 branes, this relation receives additional non-perturbative factors.
This turns out to be rather important for simplifying the computation of the 6d index as we will
discuss in section 6.5.

More generally we can consider instead of $\bC^2$ the $A_{n-1}$ orbifold as the base
of F-theory.
If we do not add any further blow ups, this gives the $A_{n-1}$, $(2,0)$ theory,
which we discuss in the next section (the above modular property turns out to be important later when we compute, in our formalism,
the index of an M5 brane).  If in addition we also blow up the points
in the base we get among the various possibilities the small $E_8$ instantons
in the $A_{n-1}$ geometry, as $(1,0)$ superconformal theories of the type
studied in \cite{Intriligator:1997dh}.

\subsection{Superconformal Index for ${\cal N}=(2,0)$ Theories}

${\cal N}=(2,0)$ theories occupy a unique place in all superconformal theories:  They
enjoy the most allowed supersymmetries in the highest possible dimension for superconformal
theories.  They are labeled by $ADE$ and correspond to type IIB in the presence
of ADE singularity.  The $A_{n-1}$  type is dual to $n$ coincident M5 branes.

The superconformal group in this case has $Sp(4)$ R-symmetry.  Let
$R_1$ and $R_2$ denote the two Cartans of $Sp(4)$ in an orthogonal basis,
where we view $R_2$ as the additional symmetry compared to the $(1,0)$ theory.
Then the superconformal index can be viewed as an extension of the $I_{1,0}$ 
by introducing the additional flavor symmetry $R_2-R_1$:
$$I_{(2,0)}={\rm Tr}(-1)^F {\bf q}_1^{J_{12}-R_1}{\bf q}_2^{J_{34}-R_1} {\bf q}^{J_{56}-R_1}{\bf Q}_m^{R_2-R_1}$$
The same reasoning as in the case of $(1,0)$ superconformal theories leads to the following picture.
The 5d theory we obtain by compactifying the $(2,0)$ theory is an ADE Yang-Mills
theory with 16 supercharges.
 Turning on the fugacity $Q_m$ corresponds to turning
on a mass $m$ for the adjoint field, where $Q_m=e^{2\pi i m}$ (for the identification
of this with $R_2-R_1$ generator of R-symmetry see \cite{Okuda:2010ke}).
In other words we can view the resulting theory as ${\cal N}=2^*$ theory in 5d.
Let $Z^{top}(t_i,;\tau,\tau_1,\tau_2,m)$ capture the BPS partition function for this 5d theory
where $t_i$ denotes the Cartan of ADE.
This partition function can be explicitly evaluated for the $A_{n-1}$ case using
the instanton calculus \cite{Nekrasov:2002qd,Nekrasov:2003rj} or the refined topological string \cite{Iqbal:2007ii} on the
periodic toric geometry \cite{Hollowood:2003cv}. The D and E should be in principle possible, either using geometric engineering or instanton calculus for ${\cal N}=2^*$.

Then to compute the index we have
$$I^{ADE}_{(2,0)}({-1/ \tau},\tau_1/\tau,\tau_2/\tau,m/\tau)=\int dt_i {Z^{top}(t_i;\tau_1,\tau_2,\tau,m)\over Z^{'top}({t_i\over \tau_1}; {-1\over \tau_1},{\tau_2
\over \tau_1},{\tau \over \tau_1},{m\over \tau_1})\cdot Z^{'top}({t_i\over \tau_2}; {\tau_1\over \tau_2},{-1\over \tau_2},{\tau \over \tau_2},{m\over \tau_2})},$$
where we have taken into account the relation between the 5d parameters and 6d parameters.
In order to gain insight into the mechanics of this computation we show how it works
for the simplest case, namely a single M5 brane, which corresponds to $A_0$ theory
and recover the result of \cite{Bhattacharya:2008zy}.  This lends support to our
general proposal and more specifically to the identification of the squashing
parameters and Kahler classes with the parameters appearing in the $6d$ superconformal index.
The case of $A_0$ theory is particularly simple because we have no integrals to perform.  In that
case the non-perturbative $Z$ we obtain is exactly the same as the perturbative one! 
This ends up being related to the modularity of the topological string partition function
on elliptic threefolds.  Moreover
we discuss the possibility that this may be the general story for all $(1,0)$ and $(2,0)$ theories in section 6.5.
We also
 show the setup for the computation for the higher $A_{n-1}$ theories in the
 refined topological vertex formalism.  We also give the expression for the index for the $A_1$ case in
 the unrefined setup as an integral over three factors of topological string amplitudes.
 
\subsection{Index for a single M5 brane}

\begin{figure}[t!]
\center
\includegraphics[width=6in]{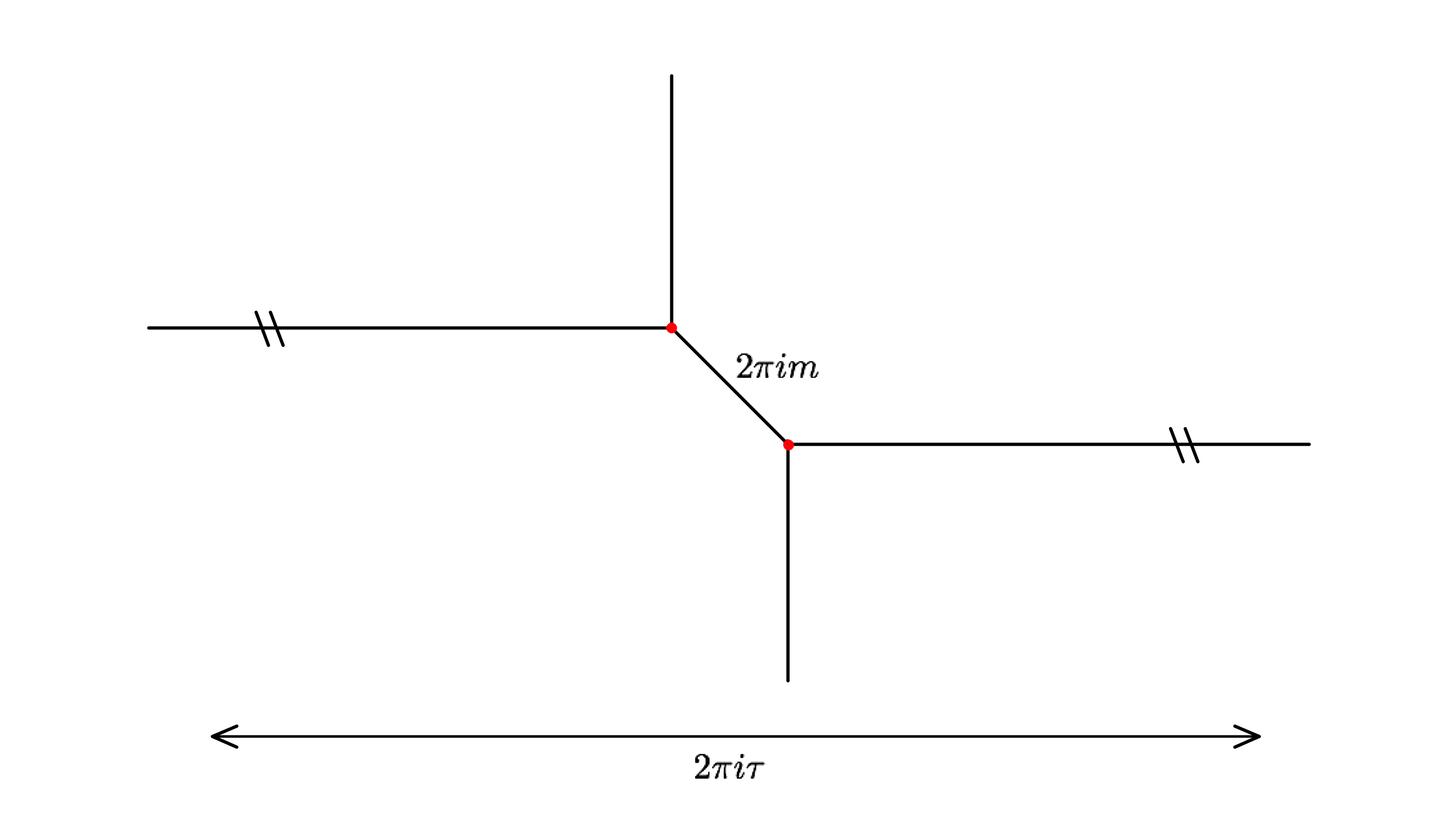}
\caption{Toric diagram for the geometry that engineers the $\scN=2^* $  $ U(1)$ theory in five dimensions. The toric plane is compactified to a cylinder, and the horizontal edges are identified with each other.}
\label{fig:u1}
\end{figure}

As discussed above the case for single M5 brane corresponds to studying topological strings
for ${\cal N}=2^*$ $U(1)$
theory in 5 dimensions.    This corresponds to a periodic toric geometry, where we compactify
the base of the toric plane along one direction, obtaining a cylinder.  The corresponding
toric diagram for this theory was introduced in \cite{Hollowood:2003cv} and extends the 4d
construction of these theories in \cite{Witten:1997sc} to 5d.  The case of $U(1)$
is shown in Figure \ref{fig:u1}.  The class corresponding to the circle identification of the
toric base is $q$ (corresponding to the elliptic fiber).  The class corresponding to the
mass parameter $m$, which we denoted by $Q_m$ is also shown in the figure.
The refined topological vertex formalism applied to this case involves introducing
the two vertices and summing over the two internal line edges with arbitrary representations,
where the smaller edge is weighted by $Q_m^n$  where $n$ is the number of boxes in the Young diagram
of the representation on that edge.  Similarly
the longer edge is weighted with $(qQ_m^{-1})^k$ where $k$ is the number of boxes in the Young
diagram of the representation on that edge.
The topological string partition function for this theory was worked out in \cite{Iqbal:2009ki} (see also \cite{nikita}) and the result is given by\footnote{We thank A. Iqbal
for a very helpful explanation of this result and its modular
properties.}(after shifting $\tau_i\rightarrow \tau_i+1$):
\[ Z_{U(1)} = \prod_{k=0}^\infty \left(\frac{\prod_{i,j=0}^\infty(1+Q_m^{-1}q^{k+1} q_1^{i+1/2}q_2^{-(j+1/2)})\prod_{i,j=0}^\infty(1+Q_mq^k q_1^{i+1/2}q_2^{-(j+1/2)})}{\prod_{i,j=0}^\infty(1-q^{k+1} q_1^{i}q_2^{-j})\prod_{i,j=0}^\infty(1-q^k q_1^{i+1}q_2^{-(j+1)})}\right),\]
where $ Q_m = e^{2\pi i m}, q = e^{2\pi i \tau}, q_1 = e^{2\pi i \tau_1}  $, and $ q_2 = e^{2\pi i \tau_2} $,
and we have included one factor of MacMahon function which is
somewhat ambiguous in the computation of the refined topological string.
The refined topological string captures the Kahler moduli dependence of the amplitudes
and does not fix the terms purely depending only on $q_1,q_2$.  In fact we will need
to multiply the above expression by $1/\eta(q_1)$ for reasons that we will explain below,
where $\eta(q_1)$ is the Dedekind eta-function.

The spectrum of this theory consists of a tower of hyper multiplets of mass $2\pi i(m+k\tau)$
(one for each integer $k$)
and a tower of tensor multiplets with mass $2\pi i k\tau$.  This is as expected, because
the reduction of a single M5 brane on a circle leads exactly to such a multiplet, where
$2\pi i \tau$ is identified with $1/R$, with $R$ the radius of the circle taking us from 6 to 5 dimensions.
It is important to rewrite the above partition function in a more symmetric way:
Let us redefine $Q_m$ by
$$Q_m\rightarrow Q_m q^{1/2}$$
Then the partition function is  totally symmetric in $(q,q_1,q_2)$, if we in addition
include a factor of $1/\eta(q_1)$ which is ambiguous
for the refined topological vertex.  To see this, we have to
rewrite everything in terms of positive powers of $q_2$:
\begin{align*}Z_{U(1)}&={1\over \eta(q_1)}\prod_{i,j,k=0}^\infty \left(\frac{(1+Q_m^{-1}q^{k+1/2} q_1^{i+1/2}q_2^{-(j+1/2)})(1+Q_mq^{k+1/2} q_1^{i+1/2}q_2^{-(j+1/2)})}{(1-q^{k+1} q_1^{i}q_2^{-j})(1-q^k q_1^{i+1}q_2^{-(j+1)})}\right)\\
&={1\over \eta(q_1)}\prod_{i,j,k=0}^\infty \left(\frac{(1-q^{k+1} q_1^{i}q_2^{j+1})(1-q^k q_1^{i+1}q_2^{j})}{(1+Q_m^{-1}q^{k+1/2} q_1^{i+1/2}q_2^{j+1/2})(1+Q_mq^{k+1/2} q_1^{i+1/2}q_2^{j+1/2})}\right)\\
&={1\over \eta(q)\eta(q_1)\eta(q_2)}\prod_{i,j,k=0}^\infty \left(\frac{(1-q^{k+1} q_1^{i+1}q_2^{j+1})(1-q^k q_1^{i}q_2^{j})}{(1+Q_m^{-1}q^{k+1/2} q_1^{i+1/2}q_2^{j+1/2})(1+Q_mq^{k+1/2} q_1^{i+1/2}q_2^{j+1/2})}\right),\end{align*}
where we delete the $i=j=k=0$
terms for the second term in the numerator.  The manifest permutation symmetry between $q,q_1,q_2$
is expected from the fact that in the 6d they become the parameters associated to the three
rotation planes.  Note also that the way we have rewritten the numerator corresponds
to the fact that a tensor multiplet in 5d is dual to the vector multiplet.  This accounts for the form of the numerator
which now gives a tower of vector multiplets.  In dualizing from tensor multiplets
to vectors we lose the zero modes associated to modes of the tensor multiplets
which corresponds to rotations in only one of the three planes (where $B_{i{\overline i}}$ has
a mode only in the $z_i$ direction). This accounts for the three $\eta$'s in the denominator. The reduction of the fields of the (2,0) theory to five dimensions has also been studied in detail in \cite{Bonetti:2012st}.
The partition function can be written elegantly in terms of double elliptic
gamma functions (see Appendix \ref{sec:multigamma} for a brief discussion of some of their properties):
$$G_2(z|a,b,c)=\prod_{i,j,k=0}^{\infty} (1-Z A^iB^jC^k)(1-Z^{-1}A^{i+1}B^{j+1}C^{k+1}),$$
where $(Z;A,B,C)={\rm exp}(2\pi i(z;a,b,c))$. We have
\begin{equation}Z_{U(1)}={1\over \eta(q)\eta(q_1)\eta(q_2)}\cdot {G_2'(0|\tau,\tau_1,\tau_2)\over G_2(m+{1\over 2}+{\tau+\tau_1+\tau_2\over 2}|\tau,\tau_1,\tau_2)}.\label{eq:zpert}\end{equation}
where we are deleting the zero mode of $G_2(0)$ as noted before. To construct the partition function of this theory on $S^5$ we simply have to consider
the above topological string partition function and take
three copies of it for the modes of the vector multiplet and the hypermultiplet on the $S^5$.  Dropping for now the factors of $ \eta $, we get:
\begin{align*}\displaystyle\dfrac{{G_2'(0|\tau,\tau_1,\tau_2)\over G_2(m-{1\over 2}+{\tau+\tau_1+\tau_2\over 2}|\tau,\tau_1,\tau_2)}}{ {G_2'(0|\tau/\tau_1,-1/\tau_1,\tau_2/\tau_1)\over G_2((m-{1\over 2}+{\tau+\tau_1+\tau_2\over 2})/\tau_1|\tau/\tau_1,-1/\tau_1,\tau_2/\tau_1)} \cdot {G_2'(0|\tau/\tau_2,\tau_1/\tau_2,-1/\tau_2)\over G_2((m-{1\over 2}+{\tau+\tau_1+\tau_2\over 2})/\tau_2|\tau/\tau_2,\tau_1/\tau_2,-1/\tau_2)}}.\end{align*}
The non-perturbative contributions to the partition function of an arbitrary 5d theory can a priori be quite complicated, but, in fact, here we find that they cancel out! This is because elliptic gamma functions satisfy a beautiful modular property \cite{Narukawa:2003}:
\begin{align}  G_2(z|\tau_0,\tau_1,\tau_2) &= \exp\left(\frac{\pi i}{12}B_{44}(z|\tau_0,\tau_1,\tau_2,1)\right) G_2\left(\frac{z}{\tau_0}\bigg|-\frac{1}{\tau_0},\frac{\tau_1}{\tau_0},\frac{\tau_2}{\tau_0}\right)\notag\\
&\qquad\qquad G_2\left(\frac{z}{\tau_1}\bigg|\frac{\tau_0}{\tau_1},-\frac{1}{\tau_1},\frac{\tau_2}{\tau_1}\right)\cdot
G_2\left(\frac{z}{\tau_2}\bigg|\frac{\tau_0}{\tau_2},\frac{\tau_1}{\tau_2},-\frac{1}{\tau_2}\right).
\label{impo}\end{align}
Using this, the expression above simplifies to\footnote{In the following manipulations we
do not keep track of the cubic and quartic prefactors which arise as a result of modular transformations. It would be interesting to understand these factors in greater detail.}
\begin{align*}{G_2'(0|-1/\tau,\tau_1/\tau,\tau_2/\tau)\over G_2((m/\tau+1/2(1+(\tau_1+\tau_2-1)/\tau)|-1/\tau,\tau_1/\tau,\tau_2/\tau)}.\end{align*}
It is remarkable that taking the three copies of the five-dimensional partition function led to an answer which is  perturbative in $ \tau_1,\tau_2$, and we offer an explanation of it below.

Likewise, the contributions from the $ \eta $ factors simplify. From $ \eta(\tau_1)\eta(\tau_2) $ we get, up to prefactor:
\[ \eta(\tau_1)\eta(\tau_2) \to \frac{\eta(\tau_1)\eta(\tau_2)}{\eta(-1/\tau_1)\eta(-1/\tau_2)\eta(\tau_1/\tau_2)\eta(\tau_2/\tau_1)} = 1.\]
From $ \eta(\tau) $ we get
\[ \frac{\eta(\tau)}{\eta(\tau/\tau_1)\eta(\tau/\tau_2)}=\eta(-1/\tau)\eta(-\tau_1/\tau)\eta(-\tau_2/\tau)=\eta(-1/\tau)\eta(\tau_1/\tau)\eta(\tau_2/\tau).\]
We thus end up with
\begin{equation}
Z_{U(1)}^{np} = \frac{1}{\eta(-1/\tau)\eta(\tau_1/\tau)\eta(\tau_2/\tau)}{G_2'(0|-1/\tau,\tau_1/\tau,\tau_2/\tau)\over G_2(\frac{m}{\tau}+\frac{1}{2}+\frac{\tau_1+
\tau_2-1}{2\tau}|-1/\tau,\tau_1/\tau,\tau_2/\tau)}\label{eq:znopert}
\end{equation}
A glance at equations \eqref{eq:zpert} and \eqref{eq:znopert} reveals that the only difference between the perturbative answer and the full non-perturbative result is a rescaling of
\[ (m,\tau,\tau_1,\tau_2) \to (m/\tau,-1/\tau,\tau_1/\tau,\tau_2/\tau),\]
which is the correct map between the 5d and 6d parameters, as discussed above. 
We now offer an explanation of the fact that the non-perturbative
completion of the $Z^{top}$ resulted in the same function in modular
transformed variables.   As discussed before (and which can be verified explicitly
for this example), we expect a pertubative modularity of the topological string partition
functions of elliptic Calabi-Yau threefold of the form:
$$Z^{top}(m_i,\tau,\tau_1,\tau_2)=Z^{top}(m_i/\tau,-1/\tau,\tau_1/ \tau, \tau_2/\tau).$$
Instead what we have found in this example is that 
$${Z^{top}(m,\tau,\tau_1,\tau_2)\over Z^{top}(m/\tau_1,\tau/\tau_1, -1/\tau_1,\tau_2/\tau_1)\cdot 
Z^{top}(m/\tau_2, \tau/\tau_2,\tau_1/\tau_2,-1/\tau_2)}
=Z^{top}(m/\tau,-1/\tau,\tau_1/ \tau, \tau_2/\tau).$$
Note that the additional terms in the denominator are non-perturbative
in the topological string coupling constants and thus can be viewed as a non-perturbative
completion of the modularity of topological strings.   We will comment on the implication of
this for possible simplification for the general computation of the index of all 6d theories in section 6.5.

The same result could also have been derived from the relation between the triple sine and elliptic gamma functions (equation \eqref{eq:s3g2}), which we also report here:
\begin{align*}G_2(z|\underline\omega)&=\exp\left(\frac{2\pi i}{4!}B_{4,4}(z|(\underline\omega,-1))\right)\notag\\
&\;\; \cdot\prod_{k=0}^\infty\frac{S_{3}(z+k+1|\underline{\omega})S_{3}(z-k|\underline{\omega})}{\exp\left(\frac{\pi i}{3!}(B_{3,3}(z+k+1|\underline\omega)-B_{3,3}(z-k|\underline\omega))\right)}.\end{align*}
Let us now denote $ e^{-2\pi i/\tau}, e^{2\pi i m/\tau}, e^{2\pi i\tau_1/\tau} $ and $ e^{2\pi i\tau_2/\tau} $ respectively by $ \mathbf{q},\mathbf{q}_m,\mathbf{q}_1,\mathbf{q}_2 $. Then, using equation \eqref{eq:plethys}, we can write
\[G_2(0|-1/\tau,\tau_1/\tau,\tau_2/\tau) = \exp\left(-\sum_n\frac{1}{n}\frac{1+\mathbf q_1^n\mathbf q_1^n\mathbf q_2^n}{(1-\mathbf q^n)(1-\mathbf q_1^n)(1-\mathbf q_2^n)}\right)\]
and
\begin{align*}&\quad G_2\left(\frac{m}{\tau}+\frac{1+\tau_1/\tau+\tau_2/\tau-1/\tau}{2}\bigg|-\frac{1}{\tau},\frac{\tau_1}{\tau},\frac{\tau_2}{\tau}\right)^{-1}\\
&\;= \exp\left(\sum_n\frac{1}{n}\frac{(\mathbf q\mathbf q_1\mathbf q_2)^{n/2}((-\mathbf q_m)^n+(-\mathbf q_m)^{-n})}{(1-\mathbf q^n)(1-\mathbf q_1^n)(1-\mathbf q_2^n)}\right).\end{align*}
Likewise,
\[ \frac{1}{\eta(-1/\tau)} = \exp\left(\sum_{k=1}^\infty\sum_{n=1}^\infty \frac{\mathbf q^{2\pi i nk}}{n} \right)=\exp\left(\sum_n\frac{1}{n}\frac{\mathbf q^n}{1-\mathbf q^n}\right),\]
and similarly for $\eta(\tau_1)$ and $\eta(\tau_2)$.   Writing
\[Z_{U(1)}^{np}=\exp\left(\sum_n\frac{I(\mathbf q_m^n,\mathbf q^n,\mathbf q_1^n,\mathbf q_2^n)}{n}\right),\]
we get
\[ I=\frac{\mathbf q}{1-\mathbf q}+\frac{\mathbf q_1}{1-\mathbf q_1}+\frac{\mathbf q_2}{1-\mathbf q_2}+\frac{\sqrt{\mathbf q\mathbf q_1\mathbf q_2}(-\mathbf q_m-\mathbf q_m^{-1})-1-\mathbf q\mathbf q_1\mathbf q_2}{(1-\mathbf q)(1-\mathbf q_1)(1-\mathbf q_2)}\]
\[ =\frac{\sqrt{\mathbf q \mathbf q_1\mathbf q_2}(-\mathbf q_m-\mathbf q_m^{-1})+\mathbf q\mathbf q_1\mathbf q_2-\mathbf q\mathbf q_1-\mathbf q\mathbf q_2-\mathbf q_1\mathbf q_2}{(1-\mathbf q)(1-\mathbf q_1)(1-\mathbf q_2)}-1.\]
Deleting the zero mode of $G_2(0)$ correspond to deleting the $-1$ in the above expression.
The resulting expression matches exactly with the result of \cite{Bhattacharya:2008zy},
\[I = \frac{x^6(z^{1/2}+z^{-1/2})+x^{12}-x^{8}(y_2+1/y_1+y_1/y_2)}{(1-x^4y_1)(1-x^4/y_2)(1-x^4y_2/y_1)},\]
provided that we identify
\[ x^4y_2/y_1 =\mathbf q,\quad x^4/y_2 =\mathbf q_2,\quad x^4y_1 = \mathbf q_1,\quad -z^{1/2}=\mathbf q_m,\]
which is in accord with the transformation of the basis used in that paper compared to ours in writing
the index.

\subsection{Multiple M5 branes}

\begin{figure}[t!]
\center
\includegraphics[width=4.5in]{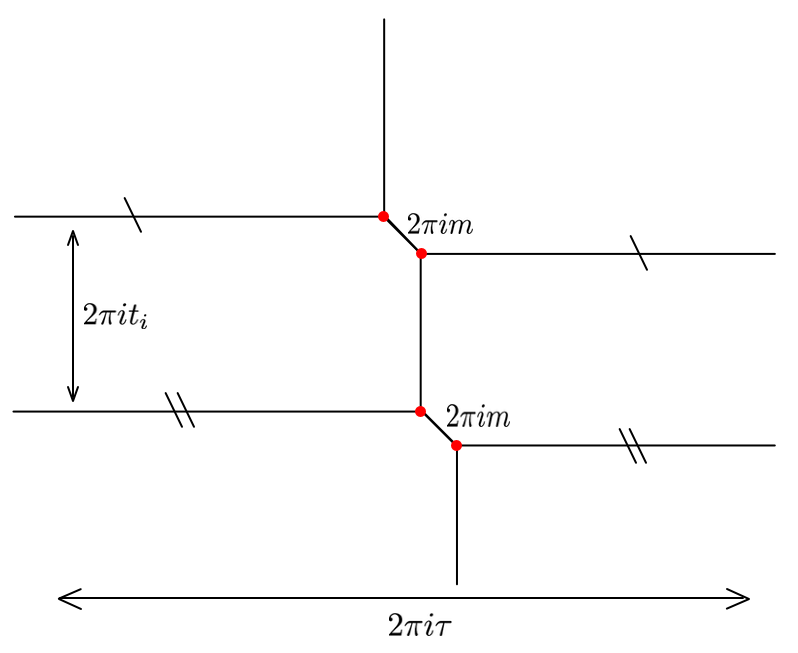}
\caption{The periodic toric geometry for the $A_1$ case.  For $A_{n-1}$ case we
get $n$ horizontal lines.  The $t_i$ are the Coulomb branch parameters and $m$ corresponds
to the mass of the adjoint in the $N=2^*$ theory.}
\label{fig:toric2}
\end{figure}

Similarly we can consider multiple M5 branes.  This was studied in \cite{Hollowood:2003cv} in
the unrefined topological string formalism  (where $q_1q_2=1$)  which can easily be generalized to the refined one (which was not developed at the time).   For $N$ M5-branes the toric geometry
will involve $N$ parallel lines wrapping the periodic direction of the toric base.   See Figure \ref{fig:toric2}
for the case with $N=2$.  The topological string will depend on one mass parameter $m$, on the periodic
size $\tau$, and on $N-1$ moduli $t_i$ which correspond to relative separation of the horizontal lines.  These
are the parameters that we need to integrate over in evaluating the 6d index.
It would be interesting to perform this computation in detail \cite{Iqbal}.  This involves gluing
$2N$ vertices of the refined topological vertex, and a sum over $3N$ Young diagrams
attached to the internal edges, just as in the unrefined case (where $\tau_1+\tau_2=0$) studied in detail in \cite{Hollowood:2003cv}. In that case, the answer for topological string partition function is given by

\begin{align*}
Z_{top}&=M(q) \prod_{k=0}^{\infty}\Big( (1-Q_{m}q^{k+1})^{2}\,
(1-Q_{F}Q_m^{-1}q^{k+1})(1-Q_{F}Q_{m})
(1-Q_{F}q^{k+1})^{-2}\Big)^{k+1}\\
&\;\;\;\cdot\sum_{\nu_1, \nu_2}\Bigg\{(QQ_{m})^{|\nu_1|+|\nu_2|}\prod_{\substack{p=1,2\\(i,j)\in \nu_p}}\frac{(1-Q_{m}q^{h(i,j)})(1-Q_{m}^{-1}q^{h(i,j)})}{(1-q^{h(i,j)})^{2}}\\
&\;\;\;\cdot\prod_{k}\Big(\frac{(1-Q_{F}Q_{m}^{-1}q^{k})(1-Q_{F}Q_{m}q^{k})
}{(1-Q_{F}q^{k})^{2}}\Big)^{C_{k}(\nu_1,\nu_2^{t})}\Bigg\},
\end{align*}
where $ M(q) $ is the MacMahon function, $q= e^{2\pi i \tau_1}, Q = e^{2\pi i \tau}, Q_{F}=e^{2\pi i a},$ and $ Q_m = e^{2\pi i m},$
where we have denoted $ e^{2\pi i \tau} $ by $ Q $ instead of $ q $ since $ q $ parametrizes the unrefined topological string coupling constant. Also, $ h(i,j) = \nu_i-i+\nu^t_j-j+1 $ is the hook length for a box $ (i,j)\in \nu $, and $ C_k(\nu_1,\nu_2) $ can be computed from
\begin{align*} \sum_{k}C_{k}(\nu_{1},\nu_{2})q^{k}=
\frac{(q-1)^2}{q}\ f_{\nu_{1}}f_{\nu_{2}}+f_{\nu_{1}}+f_{\nu_{2}}\,, \end{align*}
where  $f_{\nu}(q)=\sum_{(i,j)\in \nu}q^{j-i}$.   This can be extended to the refined
computation which we denote by $ Z_{top}(\tau,a,m;\tau_1,\tau_2) $, from which we would compute the full
index by doing the integral over the $a$ variable for the $Z_{np}$.

\subsection{$Z_{top}=Z_{np}$ in 6d?}

As we have seen in the context of computation of the superconformal index for a single M5 brane,
the non-perturbative completion of $Z_{top}$ yields again $Z_{top}$ with modular transformed 
variables.  This raises the question whether this is always true, namely\footnote{We would like
to thank D. Jafferis for discussions on this point.}:
$$Z_{np}(t_i, m_j, \tau ,\tau_1,\tau_2)=Z_{top} (t_i/\tau,m_j/\tau,-1/\tau, \tau_1/\tau,\tau_2/\tau)?$$
However, as already discussed, we expect from the perturbative modularity of $Z_{top}$ a relation of
almost this form, namely
$$Z_{top}(t_i, m_j, \tau ,\tau_1,\tau_2)=Z_{top} (t_i,m_j/\tau,-1/\tau, \tau_1/\tau,\tau_2/\tau)\bigg|_{pert.}$$
This is almost of the naive form we expected, except that $t_i$, the dynamical variables
which we need to integrate over, are not transformed under $\tau \rightarrow -1/\tau$.
This strongly suggests that the non-perturbative completion of the above equation is simply
$$Z_{np}(t_i, m_j, \tau ,\tau_1,\tau_2)=Z_{top} (t_i,m_j/\tau,-1/\tau, \tau_1/\tau,\tau_2/\tau).$$
This would be consistent with the fact that the BPS states of the elliptic 3-fold
should organize according to a tower of KK modes and for each such tower
the identity \ref{impo} would transform the answer back to the original form except
in the modular transformed variables.
This would give a dramatic simplification for the computation for the 6d case.
Namely we would get (taking into account the change of parameters from 5d to 6d):
$$I^{6d}(m_j,\tau,\tau_1,\tau_2)=\int dt_i\  Z_{top}(t_i, m_j, \tau ,\tau_1,\tau_2) $$
where $Z_{top}$ is the same as the 5d gauge theory partition function (including the cubic prefactor).
We are currently investigating this theory \cite{Haghighat}.

\subsection{Superconformal index for ${\cal N}=1,2$ in $d=4$}

Similarly in the above context we can consider the open string sectors.  These will support
4d field theories in the following way:  Consider again F-theory on elliptic 3-folds and consider
$(p,q)$ 5-branes of IIB wrapped around the Lagrangian 2-cycles of the base.  This of course needs
to be compatible with the elliptic fibration structure of F-theory as the 5-branes transform
under $SL(2,{\bf Z})$.  This leads to an ${\cal N}=1$, $d=4$ theory living on the uncompactified directions
of the 5-brane.  To the best of our knowledge these theories have not been studied before.
It would be interesting to investigate this class of theories.

 Upon further compactification on a circle, where we wrap
one of the directions of the brane on the circle,  this will correspond to a 3d theory
living on its world volume.  By the duality between F-theory and M-theory, this corresponds
to M5 branes wrapping Lagrangian cycles of the resulting 3-fold, which we can compactifiy on the
$S^3$ and compute the partition function, as already discussed for the
open string sector.  The corresponding index in the 4d theory is given by
$${\rm Tr}(-1)^F {\bf q}_1^{J_{12}-r}{\bf q}^{J_{34}-r}{\bf M}_i^{F_i}$$
where ${\bf q}$ can be identified with the modular transformed elliptic fiber parameter of the 3-fold and ${\bf q_1}$ corresponds
to the direction in which we have placed the brane and $F_i$ correspond to extra
symmetries one may have (associated to non-integrated Kahler classes and positions
of the brane). 

Similarly if the elliptic fibration of F-theory is constant the same construction
will lead to an ${\cal N}=2$ theory.  Here we will have one extra flavor symmetry
(the analog of the mass in the $N=2^*$ theory discussed before) which will play
the role of the additional parameter $t$ that one can add to the index
in the context of $N=2$ theories in $d=4$ \cite{malda}:
$${\rm Tr}(-1)^F {\bf q}_1^{J_{12}-r}{\bf q}^{J_{34}-r}{\bf t}^{R-r}M_i^{F_i}$$
  It would be interesting
to study these and explore connections with the computations already
done in the literature (see \cite{Gaiotto:2012xa} and references therein for examples
of such computations).

\section{Conclusion}
We have provided evidence that the partition function of superconformal
theories on $S^5$ and on $S^5\times S^1$ can be computed using closed
topological strings.  Similarly the partition function on $S^3$
and $S^3\times S^1$ associated to the open string theories can be computed in an analogous manner.  These computations involve in the closed string case
an $SL(3,{\bf Z})$ action involved in inverting the coupling
constants of the refined topological string, and in the open string case
an $SL(2, {\bf Z})$ transformation.  We used the connection
with the partition function computation to define what this inversion
precisely means and the regions of convergence of topological string
coupling constant.

These results complement that in \cite{Iqbal:2012} which shows how one can
use topological strings to compute associated partition functions
on $S^4\times S^1$ for closed topological strings and $S^2\times S^1$
for open ones, which does not involve the inversion of the string coupling
constant.  Thus altogether we have a unified picture where the partition functions of
a large class of superconformal theories which can be engineered
in dimensions 6, 5, 4, and 3 associated to Calabi-Yau threefolds or
Lagrangians in them can be computed using topological string data.
This leads to computation of all supersymmetric partition functions
in these dimensions on $S^d$ and $S^{d-1}\times S^1$ for the ones
that can be geometrically engineered, using topological strings.\footnote{Note that we can also
extend some of these results with the reults
of this paper to compute partition functions on $T^k\times S^{d-k}$.
For example, if we consider F-theory on elliptic threefolds, compactifications of $(1,0)$ and $(2,0)$ theories on
$T^2\times S^4$ can be computed
using the ideas sketched in this paper.  Similar examples
have been recently studied in the context of $T^2$ compactifications of the
5d superconformal theories \cite{Haghighat:2011xx}.}

The ideas in these papers suggest that the BPS states in a supersymmetric
theory (with enough supersymmetry) go a long
way in defining the superconformal fixed points they come from.
It would be very interesting to see whether this can be made into a
systematic
method for defining the full superconformal theory.

\section*{Note added}
After the completion of this paper a number of other papers appeared \cite{Imamura:2012bm, Kim:2012qf, Spiridonov:2012de} which have
some overlap with the current work.  In particular,  \cite{Imamura:2012bm} obtains triple sine functions
for partition functions on squashed $S^5$.  Also, the authors of \cite{Kim:2012qf} obtain a triple product structure
for the partition function for the special cases where the 5d CFT corresponds to gauge theories; the expressions they obtain are similar to ours. They also study the partition function for M5 branes using the 5d gauge theory, in agreement with the results of this paper.

\section*{Acknowledgments}
We are greatly indebted to Miranda Cheng, for participation at an earlier stage of this work.
We would also like to thank the SCGP for hospitality and a stimulating environment where we attended the 10th Simons Workshop on math and physics.   We have benefitted
from discussions with M. Aganagic, S. Cecotti, C. Cordova, T. Dimofte, A.Gadde, S. Gukov, B. Haghighat, J. Heckman, Y. Imamura, K. Intriligator, A. Iqbal, D. Jafferis, A. Klemm, J. Manschot, S. Minwalla, D. Morrison,
N. Nekrasov, V. Pestun, L. Rastelli, N. Seiberg and M. Zabzine.

The work of G. L. is supported in part by the Department of Energy Office of Science Graduate Fellowship Program (DOE SCGF), made possible in part by  the American Recovery and Reinvestment Act of 2009, administered by ORISE-ORAU under contract no. DE-AC05-06OR23100.
The work of C.V. is supported in part by NSF grant PHY-0244821. 

\appendix

\section{Multiple sine and multiple elliptic gamma hierarchies}
\label{hierarchies}

\subsection{Multiple sine hierarchy}
\label{sec:multisine}

In this appendix we provide the definition and relevant properties of the multiple sine and multiple elliptic gamma functions \cite{Jimbo:1996ss, Kurokawa:2003, Narukawa:2003, Nishizawa:2001}. We begin by defining the multiple zeta functions
\[ \zeta_r (z,s|\underline{\omega}) = \sum_{n_1,\dots,n_r = 0}^\infty (\vec{n}\cdot \underline\omega + z)^{-s},\]
 for $ z\in \bC $ and $ \text{Re } s > r $. We adopt the notation $ \underline\omega = (\omega_1,\dots,\omega_r) $ and $ \vec{n}\cdot\underline\omega = n_1\omega_1+\dots+n_r\omega_r $. We require that all $ \omega_i\in\bC $ lie within the same half of the complex plane. By analytic continuation the domain of definition of multiple zeta functions can be extended to $ s \in \bC $.
 
Multiple gamma functions are defined as
\[ \Gamma_r(z|\underline \omega) = \exp\left(\left.\frac{\partial}{\partial s} \zeta_r(s,z | \underline\omega)\right|_{s=0}\right),\]
which we can view as a regularized infinite product,
\[ \Gamma_r(z | \underline\omega) \sim \prod_{n_1,\dots,n_r = 0}^{\infty}(\vec{n}\cdot\underline\omega + z)^{-1}.  \]

Finally, the multiple sine is defined as
\begin{equation} S_r(z | \underline\omega)=\Gamma_r(z|\underline\omega)^{-1} \Gamma_r(|\underline\omega | -z|\underline\omega)^{(-1)^r},\end{equation}
where $ |\underline\omega| = \omega_1+\dots+\omega_r $.
Multiple sine functions can also be written as regularized products,
\begin{equation} S_r(z|\underline\omega) \sim \prod_{n_1,\dots,n_r = 0}^{\infty}(\vec{n}\cdot\underline\omega+|\underline\omega| - z)(\vec{n}\cdot\underline\omega + z)^{(-1)^{r+1}},\label{eq:S3def}\end{equation}
and enjoy a number of remarkable properties:
\begin{itemize}
\item \textbf{Analyticity:}

For $ r $ odd the multiple sine is an entire function in $ z $, with zeros at
\[ z = \vec{n}\cdot\underline \omega\qquad (n_1,\dots,n_r \geq 1),\]
coming from $ \Gamma_r(z|\underline\omega)^{-1} $, as well as zeros at 
\[ z = \vec{n}\cdot\underline \omega\qquad (n_1,\dots,n_r \leq 0),\]
coming from $ \Gamma_r(|\underline\omega|-z|\underline\omega)^{-1} $. For even $ r $, the multiple sine is meromorphic with zeros for $ (n_1,\dots,n_r \geq 1)$ and poles for $ (n_1,\dots,n_r \leq 0)$;

\item \textbf{Difference equation:} 
\begin{equation} S_r(x+\omega_i |\underline\omega) = S_{r-1}(x|\underline\omega(i))^{-1}S_r(x),\end{equation}
where $\underline\omega(i) = (\omega_1,\dots,\omega_{i-1},\omega_{i+1},\dots,\omega_r)$;

\item \textbf{Symmetries:} $S_r(z,\underline \omega)$ is invariant under permutations of the parameters $ \omega_i $. It also enjoys a reflection property:
\begin{align}S_r(z | \underline\omega) = S_r( |\underline\omega|-z | \underline\omega)^{(-1)^{r+1}};\label{eq:reflection}\end{align}

\item \textbf{Rescaling invariance:}
\begin{equation} S_r(c z | c\underline\omega) = S_r(z|\underline \omega),\end{equation}
for any $ c\in \bC $;

\item \textbf{Integral representation:}
In \cite{Narukawa:2003} it was shown that, when all $ \text{Re } \omega_j > 0$ and $ 0 < \text{Re } z < \text{Re }|\underline {\omega}| $, multiple sine functions can be expressed in terms of contour integrals. In particular, the double and triple sine functions have the following representation:
\begin{align} S_2(z|\omega_1,\omega_2) &= \exp\left(\frac{\pi i}{2} B_{2,2}(z|\underline \omega)+ \int_{\bR+i 0}\frac{d\ell}{\ell}\frac{e^{z\ell}}{(e^{\omega_1 \ell}-1)(e^{\omega_2 \ell} - 1)} \right),\\
S_3(z|\omega_1,\omega_2,\omega_3) &=  \exp\left(-\frac{\pi i}{6} B_{3,3}(z|\underline \omega)- \int_{\bR+i 0}\frac{d\ell}{\ell}\frac{e^{z\ell}}{(e^{\omega_1 \ell}-1)(e^{\omega_2 \ell} - 1)(e^{\omega_3 \ell} - 1)} \right),\quad\label{eq:triplesineint}\end{align}
where
\begin{align}
B_{2,2}(z|\omega_1,\omega_2) =&\; \frac{z^2}{\omega_1\omega_2}-\frac{\omega_1+\omega_2}{\omega_1\omega_2}z+\frac{\omega_1^2+\omega_2^2+3\omega_1\omega_2}{6\omega_1\omega_2},\label{eq:Bernoulli22}\\
B_{3,3}(z|\omega_1,\omega_2,\omega_3) =&\; \frac{z^3}{\omega_1\omega_2\omega_3}-\frac{3}{2}\frac{\omega_1+\omega_2+\omega_3}{\omega_1\omega_2\omega_3}z^2\notag\\
&+\frac{\omega_1^2+\omega_2^2+\omega_3^2+3(\omega_1\omega_2+\omega_1\omega_3+\omega_2\omega_3)}{2\omega_1\omega_2\omega_3}z\notag\\
&-\frac{(\omega_1+\omega_2+\omega_3)(\omega_1\omega_2+\omega_1\omega_3+\omega_2\omega_3)}{4\omega_1\omega_2\omega_3}
\label{eq:Bernoulli33}
\end{align}
are members of the family of multiple Bernoulli polynomials, which are defined as follows:
\begin{equation}\sum_{n=0}^\infty B_{r,n} (z | \underline{\omega}) \frac{t^n}{n!} = \frac{t^r e^{z t}}{\prod_{j=1}^r(e^{\omega_jt}-1)}.\end{equation}

\item \textbf{Factorization}:
When $ \text{Im } \omega_1/\omega_2 > 0 $, the double sine function can be written as the following infinite product \cite{Narukawa:2003}:
\begin{align} S_2(z | \omega_1,\omega_2) =& \exp\left(\frac{\pi i}{2}B_{2,2}(z|\omega_1,\omega_2)\right)\cdot\frac{\prod_{j=0}^\infty (1-e^{2\pi i (z/\omega_2+j\omega_1/\omega_2)})}{\prod_{j=0}^\infty(1-e^{2\pi i(z/\omega_1-(j+1)\omega_2/\omega_1)})}.\label{eq:s2factorized}
\end{align}

Similarly, when $ \text{Im }\omega_1/\omega_2 >0, \text{ Im } \omega_1/\omega_3 > 0, $ and $ \text{Im } \omega_3/\omega_2 >0 $, the triple sine factorizes as
\begin{align}
&S_3(z | \omega_1,\omega_2,\omega_3)\label{eq:s3factorized}\\
&\quad\;\;= \exp\left(-\frac{\pi i}{6}B_{3,3}(z|\omega_1,\omega_2,\omega_3)\right)\notag\\
&\quad\cdot\frac{\prod_{j,k=0}^\infty (1-e^{2\pi i (z/\omega_2+j\omega_1/\omega_2+k\omega_3/\omega_2)})\prod_{j,k=0}^\infty(1-e^{2\pi i (z/\omega_1-(j+1)\omega_3/\omega_1-(k+1)\omega_2/\omega_1)})}{\prod_{j,k=0}^\infty(1-e^{2\pi i (z/\omega_3+j\omega_1/\omega_3-(k+1)\omega_2/\omega_3)})}.\notag\end{align}
Similar expressions can be obtained for other regions by using the invariance of the triple sine function under exchange of $ \omega_1,\omega_2,\omega_3 $.

\end{itemize}

\subsection{Multiple elliptic gamma hierarchy}
\label{sec:multigamma}
When $ \omega_j \in \bH$, $ j=0,\dots,r $, the $ r $-th multiple elliptic gamma function is defined as
\begin{equation} G_r(z|\underline\omega) = \prod_{j_0,\dots,j_r= 0}^\infty(1-e^{2\pi i (z+j_0\omega_0+\dots+j_r\omega_r)})^{(-1)^r}\cdot(1-e^{2\pi i (|\underline \omega|-z+j_0\omega_0+\dots+j_r\omega_r)}).\label{eq:ellgammaproduct}\end{equation}
One can extend the definition to $ \omega_j\in\bC-\bR $ by repeated use of
\[ \prod_{p=0}^\infty (1-Xe^{2\pi i p\omega_j}) = \prod_{p=0}^{\infty} (1-Xe^{-2\pi i (p+1)\omega_j})^{-1}.\]
The multiple elliptic gamma function is related to the multiple sine function by the following identity, (which
was proved in \cite{Narukawa:2003}  if $ \text{Im }\omega_j > 0 $ for all $ j $, and $ 0 < \text{Im }z < \text{Im }|\underline{\omega}| $):
\begin{align}G_r(z|\underline\omega)&=\exp\left(\frac{2\pi i}{(r+2)!}B_{r+2,r+2}(z|(\underline\omega,-1))\right)\notag\\
&\;\; \cdot\prod_{k=0}^\infty\frac{S_{r+1}(z+k+1|\underline{\omega})^{(-1)^r}S_{r+1}(z-k|\underline{\omega})^{(-1)^r}}{\exp\left(\frac{\pi i}{(r+1)!}(B_{r+1,r+1}(z+k+1|\underline\omega)-B_{r+1,r+1}(z-k|\underline\omega))\right)}.\label{eq:s3g2}\end{align}
These functions have nice modular properties \cite{Narukawa:2003}. For example, if $ \text{Im }\tau_i \neq 0 $ and $ \text{Im }\tau_i/\tau_j \neq 0 $,
\begin{align}  G_2(z|\tau_0,\tau_1,\tau_2) &= \exp\left(\frac{\pi i}{12}B_{44}(z|\tau_0,\tau_1,\tau_2,1)\right) G_2\left(\frac{z}{\tau_0}\bigg|-\frac{1}{\tau_0},\frac{\tau_1}{\tau_0},\frac{\tau_2}{\tau_0}\right)\notag\\
&\qquad\qquad G_2\left(\frac{z}{\tau_1}\bigg|\frac{\tau_0}{\tau_1},-\frac{1}{\tau_1},\frac{\tau_2}{\tau_1}\right)\cdot
G_2\left(\frac{z}{\tau_2}\bigg|\frac{\tau_0}{\tau_2},\frac{\tau_1}{\tau_2},-\frac{1}{\tau_2}\right).\label{eq:g2modular}\end{align}
Similar formulas exist for $ r\neq 2 $. Multiple elliptic gamma functions also satisfy recursion relations, including
\[ G_r(z+1 |\tau_0,\dots,\tau_r) = G_r(z|\tau_0,\dots,\tau_r),\]
and
\[ G_r(z+\tau_i | \tau_0,\dots,\tau_r) = 1/G_r(z|\tau_0,\dots,\tau_{i-1},-\tau_{i},\tau_{i+1},\dots,\tau_r).\]
Furthermore, the infinite product representation of the multiple elliptic gamma function can written in the form of a plethystic exponential. The first factor of equation \eqref{eq:ellgammaproduct} can be written as
\begin{align*} &\quad\exp\left((-1)^r\sum_{j_0,\dots,j_r=0}^\infty \log(1-e^{2\pi i (z+j_0\omega_0+\dots+j_r\omega_r)}) \right)\\&=\exp\left((-1)^{r+1}\sum_{j_0,\dots,j_r=0}^\infty\sum_{n=1}^\infty\frac{e^{2\pi i n (z+j_0\omega_0+\dots+j_r\omega_r)}}{n} \right).\end{align*}
Resumming the geometric series corresponding to $ j_0,\dots,j_r $, we get
\begin{equation} \exp\left((-1)^{r+1}\sum_{n=1}^\infty \frac{I_r(q_z^n|q_0^n,\dots,q_r^n)}{n}\right),\end{equation}
where we defined $ q_i = e^{2\pi i \omega_i}$ for $ i = 0,\dots, r $, $ q_z= e^{2\pi i z}$, and
\[ I_r(q_z|q_0,\dots,q_r) = \frac{q_z}{\prod_{i=0}^r(1-q_i)}.\]
The other infinite product in equation \eqref{eq:ellgammaproduct} contributes a similar term, and we find that
\begin{equation} G_r(z|\underline\omega) = \exp\left(\sum_{n=1}^\infty \frac{(-1)^{r+1}I_r(q_z^n|q_0^n,\dots,q_r^n)-I_r(q_z^{-n}\cdot \prod_{j=0}^r q_j^n|q_0^n,\dots,q_r^n)}{n}\right).\label{eq:plethys}\end{equation}
Multiple elliptic gamma functions enjoy a number of other notable properties; we refer the reader to \cite{Narukawa:2003} for further details.

\section{Triple sine formulas for hyper and vector multiplets}
\label{sec:S31loop}

In this appendix we recast the one-loop hyper and vector multiplet contributions to the 5d partition function on unsquashed $ S^5 $ as computed in \cite{Kallen:2012va} in terms of triple sine functions.

\subsection{Hypermultiplets}
We wish to show that the one-loop partition function
\[ Z_{hyper} = \prod_{\mu\in R}\prod_{t}\left(t+3/2- i\phi_\mu\right)^{-(1+\frac{3}{2}t+\frac{1}{2}t^2)}, \]
for a hypermultiplet in the representation $ R $ of the gauge group, whose weights we denote by $ \mu $, is equal to
\[ \prod_{\mu} S_3(i\phi_\mu + 3/2 | 1, 1, 1)^{-1}. \]
From the definition of triple sine, we have
\[ S_3(z | 1, 1, 1) = \prod_{n_1,n_2,n_3\geq 0} (n_1+n_2+n_3+z)(n_1+n_2+n_3+3-z),\]
which can be expressed as a sum over a single integer
\[ S_3(z | 1, 1, 1) = \prod_{t\geq 0} [(t+z)(t+3-z)]^{t^2/2+3t/2+1}.\]
For each weight in the representation we have
\[S_3(i\phi_\mu+3/2 | 1,1,1)=\prod_{t\geq 0} (t+3/2+i\phi_\mu)^{t^2/2+3t/2+1} (t+3/2-i\phi_\mu)^{t^2/2+3t/2+1}.\]
By taking $ t \to -t  $ in the first factor, we can rewrite it as
\begin{align*} \prod_{t\leq 0} (-t+3/2+i\phi_\mu)^{t^2/2-3t/2+1} &= \prod_{t\leq -3} (-t-3/2+i\phi_\mu)^{(t+3)^2/2-3t/2 + 1)}\\
 &= \prod_{t\leq -3} (t+3/2-i\phi_\mu)^{(t^2/2+3t/2 + 1)},\end{align*}
up to a numerical phase. Here and in the following we will be cavalier about such numerical factors. Putting everything together, we have
\begin{align*} S_3(i\phi_\mu + 3/2 | 1, 1, 1) &= \prod_{t\geq 0} (t+3/2-i\phi_\mu)^{t^2/2+3t/2+1} \prod_{t\leq -3} (t+3/2-i\phi_\mu)^{(t^2/2+3t/2 + 1)}\\
 &= \prod_{\substack{t \in \bZ \\t\neq \{-1,-2\}}} (t+3/2-i\phi_\mu)^{t^2/2+3t/2+1}\end{align*}
Notice that when $ t = \{-1,-2\} $ the exponent $ t^2/2+3t/2+1 $ vanishes. So in fact we can write
\begin{align} S_3(i\phi_\mu + 3/2 | 1, 1, 1) = \prod_{t\in \bZ}(t+3/2-i\phi_\mu)^{t^2/2+3t/2+1},\label{eq:S3hyper}\end{align}
and indeed we find that
\[ Z_{hyper} = \prod_{\mu} S_3(i\phi_\mu+3/2 | 1, 1, 1)^{-1}.\]

\subsection{Vector multiplets}

We wish to show that the one-loop contribution from the vector multiplets,
\begin{align*} \left(\prod_{\beta>0} (i\phi_\beta)^2\right) \times Z_{vect} &= \prod_{\beta>0}\left((i\phi_\beta)^2\prod_{t\neq 0}(t^2-(i\phi_\beta)^2)^{t^2/2+3t/2+1}\right)\\
&=\prod_{\beta>0}\prod_{t\in\bZ}[(t+i\phi_\beta)(t-i\phi_\beta)]^{t^2/2+3t/2+1}\end{align*}
is equal to
\[\prod_{\beta>0}S_3(i\phi_\beta|1,1,1) S_3(3+i\phi_\beta|1,1,1).\]
To see this we simply shift $ i\phi_\beta $ by $ \frac{3}{2} $ in \eqref{eq:S3hyper} to get
\[ S_3(i\phi_\beta+3|1,1,1) = \prod_{t\in \bZ}(t-i\phi_\beta)^{t^2/2+3t/2+1}.\]
To get the other half of the answer we use 
\[S_3(i\phi_\beta|1,1,1) = S_3(-i\phi_\beta+3|1,1,1)=\prod_{t\in\bZ}(t+i\phi_\beta)^{t^2/2+3t/2+1},\]
so that indeed
\[ \prod_{\beta>0} S_3(i\phi_\beta|1,1,1)S_3(i\phi_\beta+3|1,1,1) = \prod_{\beta>0} \prod_{t\in\bZ}(t^2-(i\phi_\beta)^2)^{t^2/2+3t/2+1}.\]

\section{Zeros and poles of $C_{s_1,s_2}(z|\tau_1,\tau_2)$}
In the main text we defined the following generalization to the triple sine function:
\begin{align*}C_{s_1,s_2}(z|\tau_1,\tau_2)=\hspace{5in}\end{align*}
\begin{align}
\frac{\prod_{j,k=0}^\infty 
(1-(-1)^{2s_1+1}e^{2\pi i z/\tau_1}{\hat q}^{j-s_1+1/2}{\hat t}^{k-s_2+1/2})\cdot\prod_{j,k=0}^\infty (1-(-1)^{2s_1+1}e^{2\pi iz/\tau_2}{\tilde q}^{j+s_1+1/2}{\tilde t}^{k+s_2+1/2})}{\prod_{j,k=0}^\infty (1-(-1)^{2s_1+1}e^{2\pi iz}q^{j+s_1+1/2}t^{k-s_2+1/2})}.\label{eq:C}
\end{align}
We would like to express this in a form analogous to the definition of the triple sine function, equation \eqref{eq:S3def}. Assuming that this function has similar analytic properties to the triple sine function, we can read off the zeros $ \alpha_i $ and poles $ \beta_j $ of this function from its definition and express it as a regularized infinite product,
\[ C_{s_1,s_2}(z|\tau_1,\tau_2)\sim \frac{\prod_{i}(z-\alpha_i)}{\prod_{j}(z-\beta_j)},\]
which is valid up to an exponential prefactor.
In particular, from the denominator of \eqref{eq:C} we get
\[\prod_{j,k=0}^\infty (1-(-1)^{2s_1+1}e^{2\pi iz}q^{j+s_1+1/2}t^{k-s_2+1/2})\]
\[ \sim\prod_{j,k=0}^\infty\prod_{p=-\infty}^\infty (z+\tau_1(j+s_1+1/2)-\tau_2(k-s_2+1/2)+p+s_1+1/2) \]
\begin{equation} =\prod_{j,k=0}^\infty\prod_{p=-\infty}^\infty (\xi+\tau_1j-\tau_2(k+1)+p),\label{eq:factor1}\end{equation}
where
\[\xi = z+\tau_1(s_1+1/2)+\tau_2(s_2+1/2)+(s_1+1/2)). \]
Similarly, the numerator or \eqref{eq:C} contributes a factor of
\[\prod_{j,k=0}^\infty 
(1-(-1)^{2s_1+1}e^{2\pi i z/\tau_1}{\hat q}^{j-s_1+1/2}{\hat t}^{k-s_2+1/2})\]
\[ \sim\prod_{j,k=0}^\infty\prod_{p=-\infty}^\infty (z+\tau_1(p+s_1+1/2)-\tau_2(k-s_2+1/2)-(j-s_1+1/2)) \]
\begin{equation} =\prod_{j,k=0}^\infty\prod_{p=-\infty}^\infty (\xi+\tau_1p-\tau_2(k+1)-(j+1)), \label{eq:factor2}\end{equation}
as well as a factor of
\[\prod_{j,k=0}^\infty (1-(-1)^{2s_1+1}e^{2\pi iz/\tau_2}{\tilde q}^{j+s_1+1/2}{\tilde t}^{k+s_2+1/2})\]
\[ \sim\prod_{j,k=0}^\infty\prod_{p=-\infty}^\infty (z+\tau_1(j+s_1+1/2)+\tau_2(p+s_1+1/2)+k+s_2+1/2) \]
\[ =\prod_{j,k=0}^\infty\prod_{p=-\infty}^\infty (z+\tau_1(j+s_2+1/2)+\tau_2(p+s_2+1/2)+k+s_2+1/2) \]

\begin{equation} =\prod_{j,k=0}^\infty\prod_{p=-\infty}^\infty (\xi+\tau_1j+\tau_2p+k-s_1+s_2) =F_{s_1,s_2}\cdot \prod_{j,k=0}^\infty\prod_{p=-\infty}^\infty (\xi+\tau_1j+\tau_2p+k),\label{eq:factor3}\end{equation}
where in going from the second to the third line we used the fact that $ s_1=s_2 $ mod 1, and in the last line
\[ F_{s_1,s_2} = \begin{cases}1 \quad\text{if } s_1 = s_2\\\prod_{j=0}^\infty\prod_{p=-\infty}^\infty \prod_{l=1}^{s_1-s_2} (\xi+\tau_1j+\tau_2p-l)\quad\text{if }s_1>s_2\\\prod_{j=0}^\infty\prod_{p=-\infty}^\infty \prod_{l=s_1-s_2+1}^0 (\xi+\tau_1j+\tau_2p-l)^{-1}\quad\text{if }s_1<s_2 \end{cases}.\]

Dividing equation \eqref{eq:factor2} by \eqref{eq:factor1} gives
\begin{equation} \frac{\prod_{j,k,p=0}^\infty(\xi-\tau_1(p+1)-\tau_2(k+1)-(j+1))}{\prod_{j,k,p=0}^\infty (\xi+\tau_1j-\tau_2(k+1)+p)}; \label{ti}\end{equation}
further multiplying by factor \eqref{eq:factor3} gives
\begin{equation}C_{s_1,s_2}(z|\tau_1,\tau_2)\sim F_{s_1,s_2}\cdot\prod_{m,n,p=0}^\infty(\xi+\tau_1m+\tau_2p+n)(\xi-\tau_1(p+1)-\tau_2(n+1)-(m+1));\label{eq:Cprod}\end{equation}
in other words, 
\[C_{s_1,s_2}(z|\tau_1,\tau_2)\sim S_3(\xi|1,\tau_1,\tau_2)\cdot F_{s_1,s_2}.\]

\end{document}